\documentclass[twocolumn,twocolappendix,numberedappendix]{aastex63}

\usepackage{xspace}
\usepackage{amsmath}
\usepackage{framed} 
\usepackage{txfonts}
\usepackage{epstopdf}
\usepackage{rotating}
\usepackage{natbib}
\usepackage{ulem}
\usepackage{txfonts}
\usepackage{color}
\usepackage{hyperref}

\def\gcm3{\mathrm{g} / \mathrm{cm}^3}
\def\m200m{M_{\rm 200m}}

\def\gtsima{$\; \buildrel > \over \sim \;$}
\def\ltsima{$\; \buildrel < \over \sim \;$}
\def\prosima{$\; \buildrel \propto \over \sim \;$}
\def\gsim{\lower.7ex\hbox{\gtsima}}
\def\lsim{\lower.7ex\hbox{\ltsima}}
\def\simgt{\lower.7ex\hbox{\gtsima}}
\def\simlt{\lower.7ex\hbox{\ltsima}}
\def\simpr{\lower.7ex\hbox{\prosima}}

\usepackage{etoolbox}
\makeatletter
\patchcmd{\NAT@citex}
  {\@citea\NAT@hyper@{\NAT@nmfmt{\NAT@nm}\NAT@date}}
  {\@citea\NAT@nmfmt{\NAT@nm}\NAT@hyper@{\NAT@date}}
  {}
  {}
\patchcmd{\NAT@citex}
  {\@citea\NAT@hyper@{%
     \NAT@nmfmt{\NAT@nm}%
     \hyper@natlinkbreak{\NAT@aysep\NAT@spacechar}{\@citeb\@extra@b@citeb}%
     \NAT@date}}
  {\@citea\NAT@nmfmt{\NAT@nm}%
   \NAT@aysep\NAT@spacechar%
   \NAT@hyper@{\NAT@date}}
  {}
  {}
\patchcmd{\NAT@citex}
  {\@citea\NAT@hyper@{%
     \NAT@nmfmt{\NAT@nm}%
     \hyper@natlinkbreak{\NAT@spacechar\NAT@@open\if*#1*\else#1\NAT@spacechar\fi}%
       {\@citeb\@extra@b@citeb}%
     \NAT@date}}
  {\@citea\NAT@nmfmt{\NAT@nm}%
   \NAT@spacechar\NAT@@open\if*#1*\else#1\NAT@spacechar\fi%
   \NAT@hyper@{\NAT@date}}
  {}
  {}
\makeatother

\newcommand{\bbv}{{\bf v}}

\shorttitle{Hunting GW black holes with microlensing}
\shortauthors{Abrams \& Takada}

\begin{document}

\title{Hunting gravitational wave black holes with microlensing}
\author{Natasha~S.~Abrams}
\affiliation{Department of Astronomy, Department of Physics, Harvard University, Cambridge,
MA 02138, USA}
\author{Masahiro~Takada}
\affiliation{Kavli Institute for the Physics and Mathematics of the Universe
(WPI), The University of Tokyo Institutes for Advanced Study (UTIAS),
The University of Tokyo, Chiba 277-8583, Japan}

\date{\today}

\begin{abstract}
Gravitational microlensing is a powerful tool to search for
a population of invisible black holes (BHs)
in the Milky Way (MW), including isolated BHs and binary BHs at wide orbits that are complementary to gravitational wave observations. 
By monitoring highly populated regions of source stars like the MW bulge region, one can pursue microlensing events due to these BHs. 
We find that if BHs have a Salpeter-like mass function extended beyond $30M_\odot$
and a similar velocity and spatial structure to stars in the Galactic bulge and disk regions, the 
BH population  is a dominant source of the microlensing events at long timescales of the microlensing light curve $\gtrsim 100~$days.
This is due to a boosted sensitivity of the microlensing event rate to lens mass, given
as $M^2$, for such long-timescale events. 
A monitoring observation of $2 \times 10^{10}$
stars in the bulge region over 10 years with the Rubin Observatory Legacy 
Survey of Space and Time (LSST) would enable one to
find about
$6\times 10^5$
 BH microlensing events. 
We evaluate the efficiency of potential LSST cadences for characterizing the light curves of BH
microlensing and find that nearly all events of long timescales can be detected.
\end{abstract}


\section{Introduction}
\label{sec:introduction}

Since the first detection of GW150914 \citep{LIGO:16}, the LIGO/Virgo gravitational wave (GW) interferometer network
has and will continue to reveal a population of binary black hole (BBH) systems 
\citep{2019ApJ...882L..24A}.
All ten of the reported
BBH systems possess heavier masses than the mass scale of $5$--$10M_\sun$ that was previously anticipated from observations of X-ray binary systems \citep{1998ApJ...499..367B}. The lightest system of GW BBHs is GW170608 with
an inferred total mass of 18.7$M_\odot$
\citep{2017ApJ...851L..35A}, while the 
heaviest system is likely GW170729 \citep{2019PhRvX...9c1040A} 
with an inferred total mass of $85M_\odot$. 
Thus, the discovery of GW BBH systems has triggered an intense debate in the origin and formation channels of such massive BHs. 

There are several BBH formation channels that have been proposed in the literature. These include the following: 
from binary systems of 
massive stars, e.g. through common envelope evolution in low-metallicity environments \citep[e.g.][]{1998ApJ...506..780B,2002ApJ...572..407B,2016Natur.534..512B}; 
through mergers of lighter mass BHs in star clusters \citep[e.g.][]{2000ApJ...528L..17P}  
and galactic nuclei \citep[e.g.][]{2012ApJ...757...27A}; 
or through gas drag and stellar scattering in accretion disks surrounding super-massive BHs at the center of galaxies 
\citep[e.g.][]{2012MNRAS.425..460M,2019PhRvL.123r1101Y}.
Finally, GW BBHs 
could originate from a primordial BH population that might have formed in the early universe \citep{Sasakietal:16,Birdetal:16,2020arXiv200109160K,2020arXiv200212778C}. Thus the origin of GW BBHs 
involves rich physical processes 
and an observational exploration of BHs is mandatory in the next decade of astronomy. 

Gravitational microlensing \citep{Paczynski:86,Griestetal:91,1991ApJ...374L..37M}
can serve as a powerful tool to observationally search for BH candidates or more generally any compact objects 
in our Milky Way (MW) Galaxy.
When a lens is almost perfectly aligned with a background source star along the line-of-sight direction of an observer, 
the light of a source star is magnified, causing a characteristic light curve as a function of observing time. The timescale of a light curve depends on the lens mass and the relative velocity between 
the lens, 
source, and 
observer \citep{1995ApJ...447...53H,1996ApJ...467..540H}. 
Various experiments/observations have shown that microlensing can be used  to constrain a population of BHs and other invisible or very faint objects 
such as exoplanets, brown dwarfs and free-floating planets  \citep{Alcocketal:00,2002MNRAS.329..349M,2002ApJ...579..639B,Sumietal:03,2006Natur.439..437B,EROS:07,Sumietal:11,2017Natur.548..183M,2017arXiv170102151N,2019PhRvD..99h3503N,2020MNRAS.493.3632S,2020A&A...636A..20W}.

Hence the purpose of this paper is to study how microlensing can be used to 
study
a population of BHs in the MW Galaxy. To do this, we 
consider a scenario 
in which BHs corresponding to GW counterparts have formed from massive main-sequence  (MS) star progenitors that formed in the assembly history of our Galaxy  \citep{2020arXiv200204340H}.  
Assuming a shape of the BH mass function, 
i.e. a power-law, Gaussian, and their superposition, 
we study the expected number of microlensing events and the distribution of microlensing light-curve timescales for each of the assumed BH mass functions. The critical assumption we employ in this paper to normalize the BH mass function 
is the ``number conservation'' between MS star progenitors and the resulting BHs. 
With this assumption, we can infer  the expected number and spatial distribution of BHs from the assumed initial mass function of zero-age MS stars, if we employ
the standard model of the Galactic bulge and disk \citep{2008gady.book.....B}, which determines the abundance and spatial distribution of surviving low-mass MS stars  \citep[see][for the similar approach]{2019PhRvD..99h3503N}. 
In addition, 
we assume that BHs follow the same velocity distribution as that of MS stars in the disk and bulge regions,
which are well constrained by various observations such as the proper motion measurements by the SDSS \citep{2010ApJ...716....1B}
and {\it Gaia} datasets \citep{2018A&A...616A...1G}. 
We will then examine how different shapes of BH mass function lead to different distributions  of microlensing light curve timescales.  
To 
evaluate a concrete prospect of a BH 
search with microlensing, we will consider a hypothetical 10-year monitoring observation of stars in the Galactic bulge region 
with the Rubin Observatory Legacy Survey of Space and Time (LSST)\footnote{\url{https://www.lsst.org}}
\citep{Ivezicetal:08,2019RNAAS...3...58L,2020ApJ...889...31L,2019arXiv190201055D,2020arXiv200414347G}.

This paper is structured as follows. In Section~\ref{sec:mirolensing} we review the basics of microlensing and the event rates of microlensing for source stars in the Galactic bulge. In Section~\ref{sec:BH_massfunction} we describe the details of our model of the mass function, spatial distribution and velocity distribution of BHs. In Section~\ref{sec:results} we show the main results of this paper. 
We end with a discussion and conclusion (Section~\ref{sec:conclusion}).

\section{Microlensing basics and the event rate}
\label{sec:mirolensing}

In this section we describe our model of microlensing effects on a source star in the Galactic bulge region that are caused by lensing objects, 
stars and stellar remnants, in 
the Galactic bulge and disk regions. 

\subsection{A timescale of microlensing light curve}

When a source star and a lens are almost perfectly aligned along the line-of-sight direction of an observer, the star is multiply imaged due to strong lensing \citep{Paczynski:86}.
If these multiple images are unresolved, the flux from the star appears magnified. The light curve, or amplification, of such a microlensing magnification event is given by
\begin{align}
& A(t)=\frac{u^2+2}{u\sqrt{u^2+4}},
\label{eq:magnification}
\end{align}
where $u(t)$ is the separation between the source star and the lens at an observation epoch $t$. The impact parameter as a function of time is given by
\begin{align}
    u(t) = \sqrt{u_{\rm min}^2 + \frac{(t - t_0)^2}{t_E^2}},
\label{eq:impact_parameter}
\end{align}
where $u_{\rm min}$ is the minimum impact parameter, 
$t_0$ is the time where the two objects are closest on the sky and $t_{E}$ is the crossing time of the Einstein radius. Throughout this paper we 
use $t_{{E}}$ to characterize 
the timescale of a microlensing light curve: 
\begin{align}
t_{E}
\equiv \frac{R_{E}}{v}=\frac{\sqrt{4GM d_{\rm l}d_{\rm ls}/d_{\rm s}}}{cv},
\label{eq:omega}
\end{align}
where $R_{E}$ is the Einstein radius, $M$ is the mass of the lensing object (which is assumed to be a point mass throughout
this paper), $v$ is the (total) relative velocity on the two-dimensional 
plane perpendicular to the line-of-sight direction (see below), 
$d_{\rm l}$ and $d_{\rm s}$ are distances to 
the lens and source, respectively, and $d_{\rm ls}$ is the distance between lens and source ($d_{\rm ls}\equiv 
d_{\rm s}-d_{\rm l}$).
Note that since we are interested in massive BHs with mass $\gtrsim 10M_\odot$, we can safely 
ignore the finite source size effect and the wave effect of optical wavelengths for such massive BH microlensing \citep{2020MNRAS.493.3632S}.

If we plug typical values of the physical quantities into Eq.~(\ref{eq:omega}), we can find a typical timescale of the microlensing 
light curve as
\begin{align}
&t_{E}\simeq 242~{\rm days}\left(\frac{M}{30~M_\odot}\right)^{1/2}
\left(\frac{d_{\rm l}d_{\rm ls}/d_{\rm s}}{4~{\rm kpc}}\right)^{1/2}\left(\frac{v}{220~{\rm km/s}}\right)^{-1}.
\label{eq:t_E}
\end{align}
This equation shows that a LIGO-counterpart BH of $30~M_\odot$ mass scale causes a microlensing event whose light curve has a typical timescale of 240~days. Hence, to hunt such BHs with microlensing, we need a long 
baseline of a monitoring observation that 
spans over more than a few years. More exactly speaking, we need to take into account the distribution of relative velocity $v$ in the above equation, which causes a wider timescale distribution of the resulting microlensing light curves, as we will show below.

\subsection{Coordinate system}
\label{subsec:coordinates}

Let us first begin by defining our Cartesian coordinate 
system. We assume that the $xy$-plane is in the plane of the Galactic disk, and the $z$-direction is in the perpendicular direction to the Galactic disk. Then we take the Galactic center to be the coordinate origin and the $x$-axis to be along the direction connecting the Galactic center and the Sun. Hence we assume that the position of the Sun is given
by $(x_\odot,y_\odot,z_\odot)=(8~{\rm kpc},0,0)$. 
In this paper, we consider microlensing observations towards the Galactic bulge region since many source stars are available and the optical depth of microlensing is high, as shown in the previous observations, such as the Optical Gravitational Lensing Experiment (OGLE) \citep{1994ApJ...426L..69U,2018arXiv181100441M}. In the Galactic coordinate system ($\ell,b$), the Galactic bulge 
region is around $\ell=b=0$. For a given observation region with Galactic coordinates $(\ell,b)$, the Cartesian coordinates for a lensing object at distance $d_{\rm l}$ is given as 
$(x,y,z)=(d_{\rm l}\cos b\cos\ell, d_{\rm l}\cos b\sin\ell, d_{\rm l}\sin b)$. Following the method in \citet{2019PhRvD..99h3503N}, we consider 
$(\ell,b) =  (1.0879^\circ, -2.389^\circ)$ for a hypothetical observation region on the sky.

\subsection{Event rate of microlensing}
\label{subsec:eventrate}

The optical depth of microlensing, $\tau$, gives a probability that a {\it single} source star in the bulge region experiences microlensing 
by foreground lensing objects at a given moment. 
It is given by the line-of-sight integration of the microlensing cross section as
\begin{align}
\tau\equiv \int_0^{d_{\rm s}}\!\mathrm{d}d_{\rm l}\int\!\mathrm{d}M~ n_{\rm l}(d_{\rm l}; M)\pi R_E^2,
\end{align}
where 
$n_{\rm l}(M; d_{\rm l})\mathrm{d}M$ is the number density distribution of lenses in the mass range $[M,M+\mathrm{d}M]$ and 
at distance $d_{\rm l}$
-- more explicitly, at the coordinate position 
$(x_{\rm l},y_{\rm l},z_{\rm l})\approx (d_{\rm l},0,0)$ -- for a microlensing observation towards the Galactic bulge region. 
The dimension of $n_{\rm l}(M; d_{\rm l})\mathrm{d}M$ is
$[({\rm pc})^{-3}]$. 
Here we define the optical depth by microlensing events during which a source star and a lensing object are 
closer than the Einstein radius $R_{E}$ on the sky, or equivalently, by microlensing events with magnification greater than $A(u=R_E)\simeq 1.34$ (Eq.~\ref{eq:magnification}).
The standard Galactic model, which describes the spatial distribution of main-sequence stars in the Galactic bulge and disk regions,
yields  $\tau\sim 10^{-6}$ 
\citep{Paczynski:86,Griestetal:91,1995ApJ...447...53H}. 
That is, if we observe a million stars at once, at least one source star is expected to 
undergo
microlensing at each moment. 
For microlensing due to a lensing object in the Galactic bulge, i.e. when a lensing object and a source star are both in the Galactic bulge, 
we 
need to take into account the relative distributions of lens and 
source objects within the bulge region. 
This introduces an additional integration in the above equation. It is straightforward to 
evaluate the bulge contribution, e.g. following the method in 
\citet{2019PhRvD..99h3503N}. For notational simplicity, we give the equations for the microlensing quantities for a case that a lensing object is in the Galactic disk region.
For microlensing events of long timescales, which are 
the main focus of this paper, microlensing due to lenses in the Galactic disk 
makes a dominant contribution compared to the bulge-bulge lensing.

Now we consider the microlensing event rate for a {\it single} source star per unit observation time. 
Following the formulation in \citet{2019PhRvD..99h3503N} \citep[also see][for the pioneer work]{Griestetal:91,1995ApJ...447...53H,1996ApJ...467..540H}, 
the differential event rate of microlensing events that have a light curve timescale of $[t_{E},t_{E}+\mathrm{d}t_{E}]$
is given as
\begin{align}
\frac{\mathrm{d}\Gamma}{\mathrm{d}t_{E}}&\equiv \frac{\mathrm{d}^2\tau}{\mathrm{d}t_{\rm obs}\mathrm{d}t_E}\nonumber\\
&=2\pi
\int_0^{d_{\rm s}}\!\mathrm{d}d_{\rm l}
\int\!\mathrm{d}M~n_{\rm l}(d_{\rm l}; M)R_{E}(M, d_{\rm s},d_{\rm l})\nonumber\\
&\hspace{2em}\times \int_0^\infty\!\mathrm{d}v_\perp~\int_{-\pi/2}^{\pi/2}\!\mathrm{d}\theta~v_\perp^2\cos\theta f(v_\perp,\theta)\nonumber\\
&\hspace{2em}\times \delta_D\left(t_{E}-\frac{2R_{E}}{v_\perp \cos\theta}\right)\nonumber\\
&= \pi 
\int_0^{d_{\rm s}}\!\mathrm{d}d_{\rm l}
\int\!\mathrm{d}M~n_{\rm l}(M;d_{\rm l})\int_{-\pi/2}^{\pi/2}\!\mathrm{d}\theta~ v_\perp^4 f(v_\perp,\theta),
\label{eq:dGamma/dt_E}
\end{align}
where $v_\perp=2R_{E}/[t_{E}\cos\theta]$, 
$\bbv_\perp$ is the relative velocity vector between an observer, lens, and source star (see below) in the two-dimensional plane perpendicular to the line-of-sight direction; $\theta$ is defined via
$\bbv_\perp \equiv (v_y,v_z)= v_\perp (\cos\theta,\sin\theta)$; and  
$f(\bbv_\perp)$ is the velocity density function whose dimension is defined so that $v_\perp^2f(\bbv_\perp)$ is dimension-less. 
The dimension of the differential event rate is $[\mathrm{d}\Gamma/\mathrm{d}t_{E}]=[{\rm events/day/day}]$.
For bulge-bulge lensing, we need to perform an additional integration to take into account the relative distribution between a lens
and a source star, which is straightforward to do, e.g. following Eq.~(13) in \citet{2019PhRvD..99h3503N}.

The relative velocity for a source-lens-observer system, relevant for the microlensing event rate, is given as
\begin{align}
\bbv_\perp=\bbv_{\rm l}-\left(\frac{d_{\rm l}}{d_{\rm s}}\bbv_{\rm s}+\frac{d_{\rm ls}}{d_{\rm s}}\bbv_{\rm o}\right)
= \bbv_{\rm l}-\left[\beta\bbv_{\rm s}+(1-\beta)\bbv_{\rm o}\right],
\label{eq:v_relative}
\end{align}
where $\beta\equiv d_{\rm l}/d_{\rm s}$, and 
$\bbv_{\rm s}$, $\bbv_{\rm l}$, and $\bbv_{\rm o}$ are respectively the velocities of a source star, lens and observer (us) with respect to the rest frame of the Galactic center. We assume that the velocity of an observer is the same as that of the Sun, i.e. we ignore the motion of the Earth. That is a good approximation because the orbital motion of the Earth with respect to the Sun ($\sim 30~{\rm km~s}^{-1}$) is much smaller than that of the Sun with respect to the Galactic center ($\sim 220~{\rm km~s}^{-1}$). Throughout this paper, we assume that the velocity of an observer is 
given as $\bbv_{\rm o}=(0,220,0)~{\rm km~s}^{-1}$; that is, we assume $|\bbv_{\rm rot}|=220~{\rm km~s}^{-1}$ for  the Galactic rotation velocity. 

To gain some insight into the following results, we here discuss asymptotic behaviors of the microlensing event rate. 
In the standard Galactic models, the disk rotation of the Sun (i.e. an observer) 
is responsible for the dominant contribution to the relative velocity. 
Qualitatively, the velocity function is given as 
\begin{align}
f(\bbv_\perp) \sim \frac{1}{2\pi \sigma_v^2} \exp\left[-\frac{(\bbv_\perp -\bar{\bbv}_{\rm rot})^2}{2\sigma_v^2}\right],
\label{eq:fv_approx}
\end{align}
where $\sigma_{\rm v}$ is the velocity dispersion around the mean motion. 
Combining Eqs.~(\ref{eq:dGamma/dt_E}) and (\ref{eq:fv_approx}), we can find that the event rate for  a {\it fixed} timescale $t_{E}$
and lensing objects of mass scale $M$ is given as
\begin{align}
\frac{\mathrm{d}\Gamma}{\mathrm{d}t_{E}}&
\propto \int_0^{d_{\rm s}}\!\mathrm{d}d_{\rm l}\int\!\mathrm{d}M~n_{\rm l}(M) \frac{R_{E}^4}{t_{E}^4}
f\left(v_\perp\simeq \frac{R_{E}}{t_{E}}; v_{\rm rot},\sigma_v\right)\nonumber\\
&\propto \int_0^{d_{\rm s}}\!\mathrm{d}d_{\rm l}\int\!\mathrm{d}M~n_{\rm l}(M) \frac{M^2}{t_{E}^4}
f\left(v_\perp\simeq \frac{R_{E}}{t_{E}}; v_{\rm rot},\sigma_v\right),
\label{eq:gamma_mass_scaling}
\end{align}
where we have used the fact $R_E\propto M^{1/2}$. 
Thus we find $\mathrm{d}\Gamma/\mathrm{d}t_{E}\propto (M^2/t_E^4)\times 
f(v_\perp \sim M^{1/2}/t_{E})$. 
This means that for fixed $t_E$ the event rate is boosted for more massive lenses, assuming all lensing objects have similar spatial and velocity distributions independently of mass \citep[also see][for the earlier works]{1996ApJ...473...57M,2005MNRAS.362..945W}.
Recalling that the velocity function peaks at $v_\perp\sim v_{\rm rot}$ (Eq.~\ref{eq:fv_approx}), 
the event rate 
peaks at a particular time scale $t_E$ satisfying $t_E\sim R_E/v_{\rm rot}\propto M^{1/2}/v_{\rm rot}$. Thus, a heavier lens tends to produce a longer timescale lensing event, as expected. If we consider an even longer timescale satisfying $t_E\gg R_{E}/v_{\rm rot}$ for the same 
lens population of a fixed $M$, we find $\mathrm{d}\Gamma/\mathrm{d}t_E\propto M^2/t_E^4$ because
$f(v_\perp\simeq R_E/t_E)\sim \exp(-v_{\rm rot}^2/2\sigma_v^2)
\sim {\rm constant}$ for $v_\perp\simeq R_E/t_E\ll v_{\rm rot}$; that is, the event rate has 
similar dependencies on $t_E$ and $M$
for all lens populations at this limit.
With these facts in mind, we can see
that, even if an abundance of BHs of $30M_\odot$ mass and above
is much smaller than that of main-sequence stars at $M\simlt 1~M_\odot$, BHs can 
be a dominant source of long timescale microlensing 
events, as we will show below more quantitatively.

Throughout this paper we ignore cases 
in which the microlensing source star is in the disk region, i.e. 
disk-disk microlensing. 
An actual observation would avoid the very center of the Galactic bulge region due to the 
high dust extinction.
Instead, a survey would likely target 
regions a few degrees away
from the Galactic center in 
the Galactic latitude 
and/or longitude directions.
Since the optical depth for 
disk-disk microlensing rapidly decreases with the angular distance from the Galactic center as shown in \citet{2020ApJS..249...16M}, we do not think that 
ignoring the disk-disk lensing 
largely changes the following results. However, in
disk-disk microlensing
the lens and 
source 
tend to have a small relative velocity since they are both in the Galactic rotation motion, so it would lead to a longer timescale 
microlensing event, even if the lens 
is of solar mass scale or smaller (a typical main-sequence star mass) \citep[also see][for a similar discussion]{2016MNRAS.458.3012W}. For this case, we believe that the 
microlensing parallax can be used to discriminate such an event from a BH-lens event, and 
we will come back to this possibility later.

\section{Mass function and spatial and velocity distributions for BH}
\label{sec:BH_massfunction}

To evaluate the microlensing event rate,
we need to model 
the number density distribution (spatial distribution) and the velocity distribution of lensing objects in the Galactic bulge and disk regions. In this section 
we describe our models of these quantities. 

\subsection{Mass functions of ZAMS stars and stellar remnants}
\begin{figure}
\centering
\includegraphics[width=0.48\textwidth]{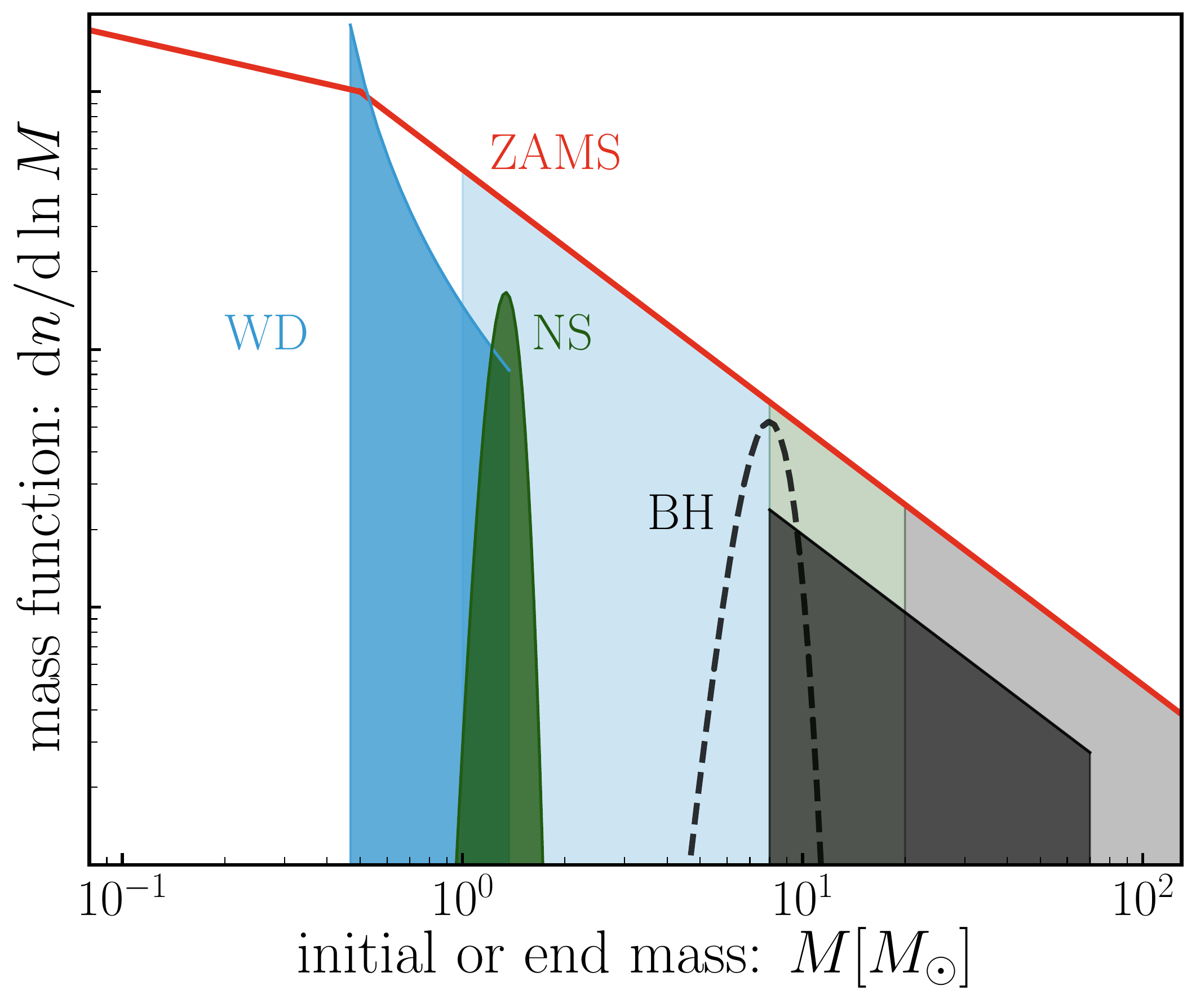}
\caption{Broken power-law, red curve denotes the initial mass function of main-sequence stars assuming 
the Kroupa-like model (see Eq.~\ref{eq:IMF}).
Note that $y$ axis is in an arbitrary scale. We assume that each massive star with $M_{\rm ZAMS}\ge 1~M_\odot$ has already evolved 
into its respective stellar remnant: 
white dwarfs (WD) for stars with $1\le M_{\rm ZAMS}/M_\odot\le 8$ following the initial and end mass relation, 
$M_{\rm WD}=0.339+0.129M_{\rm init}$ (blue line), neutron stars (NS) for $8\le M_{\rm ZAMS}/M_\odot \le 20$ (green), and 
stellar origin black holes (BH)
for $M_{\rm ZAMS}\ge 20~M_\odot$, respectively. For BHs, we consider two models of the mass function; the black dashed line shows 
a Gaussian model with mean $M_{\rm final}=7.8M_\odot$ and width
$\sigma=1.2M_\odot$, while the black solid line shows the power-law mass function with slope $\alpha_{\rm BH}=2$, i.e. the same slope as
that for the ZAMS stars at the high mass end.
For NS,
we adopt
the Gaussian model
with 
$M_{\rm final}=1.33M_\odot$ and $\sigma=0.12M_\odot$ (green line).
Because of the number conservation, the area under the curve for each stellar remnant, $\int\mathrm{d}\ln M~\mathrm{d}n/\mathrm{d}\ln M$, is the same as the area of the IMF over the corresponding range of initial main-sequence star masses (the two shaded regions of similar color have 
the same area). 
}
\label{fig:IMF}
\end{figure} 

We need to model mass functions (abundances) of lensing stars and stellar remnants to calculate the microlensing event rate. 
By stellar remnants we mean white dwarfs (WD), neutron stars (NS) and BHs, which are very faint or invisible, and therefore their genuine abundances 
are poorly understood.  
Throughout this paper we adopt the unified initial mass function of stars at birth, i.e. zero-age main-sequence stars (ZAMS), in both the Galactic bulge and 
disk regions. We then assume that, by today, massive stars of $M_{\rm ZAMS}\ge M_\odot$ have already evolved to form stellar remnants (WD, NS and BH).  
For visible objects, i.e. surviving main-sequence stars, we have a good idea of their spatial distribution in both the 
Galactic bulge and disk regions, which is hereafter referred to as the standard model of the 
Galactic 
structure.
Our approach is to estimate the abundance of each stellar remnant population relative to that of main-sequence (MS) stars, assuming the number conservation between the ZAMS progenitors of the remnant population at birth and the remnants today.
In our method, we include neither possible spatial inhomogeneities in the ZAMS mass function 
or 
stellar populations nor
the star formation history, 
for simplicity. 
These simplified assumptions are appropriate here because we consider the abundances of stellar remnants relative to that of MS stars observed today. In addition, as long as the initial mass function of stars at birth has a similar form at each epoch
of the star formation activities in the past (extending to sufficiently massive stars), the following results capture the main characteristics of the microlensing event rates. 
Nevertheless, these assumptions need to be revisited more carefully, and we should keep them in mind as a caveat of this approach. 

We assume the following broken power-law mass function for ZAMS stars at birth:  the number of stars in each logarithmic mass interval is given as 
\begin{align}
&\frac{\mathrm{d} n_{\rm s}(M)}{\mathrm{d}\ln M}=\left\{
\begin{array}{ll}
{\displaystyle A_{\rm MS}\left(\frac{M}{0.5~M_\odot}\right)^{1-\alpha_{\rm MS1}} }\, , & (0.08\le M/M_\odot \le 0.5)\\
{\displaystyle A_{\rm MS}\left(\frac{M}{0.5~M_\odot}\right)^{1-\alpha_{\rm MS2}} }\, , & (M/M_\odot \ge 0.5),
\end{array} 
\right. 
\label{eq:IMF}
\end{align}
where $A_{\rm MS}$ is the normalization constant that we will determine later; 
$A_{\rm MS}$ has a dimension of $[({\rm pc})^{-3}]$. In the following, we often omit the subscript ``ZAMS'' to denote mass of ZAMS star, $M_{\rm ZAMS}$, 
for notational simplicity.
We adopt the Kroupa-like model for the mass function. Our default choices are
 $\alpha_{\rm MS1}=1.3$ and $\alpha_{\rm MS2}=2.0$ 
\citep{2001MNRAS.322..231K}, because such a model nicely reproduces the timescale distribution of microlensing events in the OGLE observation 
as shown in \citet{2018arXiv181100441M}
\citep[also see][]{Sumietal:03,Sumietal:11,2019PhRvD..99h3503N}. 
The number of ZAMS stars is dominated by low-mass stars.

Massive stars with $M\ge M_\odot$ have rapid evolution, and we assume that such massive stars have already evolved to stellar remnants. We assume 
ZAMS stars with $M_\odot\le M_{\rm ZAMS}\le 8M_\odot$ have evolved to WDs,
ZAMS stars with $8M_\odot \le M_{\rm ZAMS}\le 20M_\odot$ to NSs,
and ZAMS stars with $20M_\odot \le M_{\rm ZAMS}\le 100M_{\odot}$ to BHs.
The critical assumption that we adopt throughout this paper is the number conservation between ZAMS stars and the stellar remnants. That is,
we impose the number conservation: 
\begin{align}
\int_{M^{\rm ZAMS}_{{\rm SR},b}}^{M^{\rm ZAMS}_{{\rm SR},u}}\!\mathrm{d}\ln M\frac{\mathrm{d}n_{\rm s}}{\mathrm{d}\ln M}=
\int_{0}^{\infty}\!\mathrm{d}\ln M_{\rm SR}~\phi(M_{\rm SR}),
\end{align}
where $\phi(M_{\rm SR})$ denotes the mass function of each stellar remnant (SR) population (WD, NS or BH) in the logarithmic mass interval, 
and
$M^{\rm ZAMS}_{{\rm SR},b}$ and $M^{\rm ZAMS}_{{\rm SR},u}$ denote the lower- and upper-boundary masses,
respectively, for 
the progenitor ZAMS stars of each stellar remnant; e.g., for WD, $M^{\rm ZAMS}_{{\rm WD},b}=M_\odot$ and 
$M^{\rm ZAMS}_{{\rm WD},u}=8M_\odot$. $M_{\rm SR}$ is the mass of the stellar remnant.

In the following, we describe the mass function for each population and how to determine the normalization. 
\begin{itemize}
	\item White dwarf (WD) -- We assume that WDs form from ZAMS stars whose 
		mass is not high enough to have a supernova explosion ($1\le M_{\rm ZAMS}/M_\odot\le 8$). 
	After the red giant stage in the stellar evolutionary track, WDs are formed from the core of a star composed of carbon and oxygen. Here we adopt a simple mass conversion between the progenitor ZAMS mass and the final WD: $M_{\rm WD}=0.339+0.129M$ 
	($M_{\rm WD}$ and $M$ are in units of $M_\odot$) \citep{2009ApJ...693..355W}. 
	Hence the mass function of WD is computed from 
	\begin{align}
	\phi_{\rm WD}(M_{\rm WD})\mathrm{d}\ln M_{\rm WD}=\left.\frac{\mathrm{d}n_{\rm s}}{\mathrm{d}\ln M}\right|_{M=M(M_{\rm WD})}\mathrm{d}\ln M .
	\end{align}
	The WD mass function is found to be
	\begin{align}
	\phi_{\rm WD}(M_{\rm WD})&= \left.\frac{\mathrm{d}n_{\rm s}}{\mathrm{d}\ln M}\right|_{M=M(M_{\rm WD})}\frac{\mathrm{d}\ln M}{\mathrm{d}\ln M_{\rm WD}} \nonumber\\
		&\hspace{-3em}= A_{\rm MS}\left(\frac{M_{\rm WD}-0.339}{0.129\times 0.5M_\odot}\right)^{1-\alpha_{{\rm MS}2}}
		\frac{M_{\rm WD}}{M_{\rm WD}-0.339}.
		\label{eq:phi_WD}
	\end{align}

	\item Neutron star (NS) -- NSs originate from core-collapse supernovae of massive stars. Here we assume that massive ZAMS stars with $8\le M_{\rm ZAMS}/M_\odot\le 20$ evolve into NSs. We employ a simplified model; we assume that the mass function of NS follows a Gaussian distribution, so number conservation gives
	\begin{align}
	\int_{0}^{\infty}\mathrm{d}\ln M_{\rm NS} \phi_{\rm NS}(M_{\rm NS})=\int_{M=8}^{M=20}\mathrm{d}\ln M \frac{\mathrm{d}n_{\rm s}}{\mathrm{d}\ln M},
	\label{eq:ns_norm}
	\end{align}
	with
	\begin{align}
	\phi_{\rm NS}(M_{\rm NS})\mathrm{d}\ln M_{\rm NS}\equiv \frac{A_{\rm NS}}{\sqrt{2\pi}\sigma_{\rm NS}}\exp\left[-\frac{(M_{\rm NS}-\bar{M}_{\rm NS})^2}{2\sigma_{\rm NS}^2}\right]\mathrm{d}M_{\rm NS},
	\label{eq:phi_NS}
	\end{align}
	where $A_{\rm NS}$ is the normalization parameter (see below), and we adopt $\bar{M}_{\rm NS}=1.33M_\odot$ and $\sigma_{\rm NS}=0.12M_\odot$. 
	Eq.~(\ref{eq:ns_norm}) relates the normalization parameter $A_{\rm NS}$ to the normalization parameter of ZAMS, $A_{\rm MS}$, as 
	\begin{align}
	A_{\rm NS}=\frac{A_{\rm MS}}{1-\alpha_{\rm MS2}}\left[
	\left(\frac{20}{0.5}\right)^{1-\alpha_{\rm MS2}}-	\left(\frac{8}{0.5}\right)^{1-\alpha_{\rm MS2}}
	\right].
	\label{eq:A_NS}
	\end{align}
	Once the normalization of ZAMS mass function, $A_{\rm MS}$, is given, it determines the normalization of the NS mass function, $A_{\rm NS}$.
	Due to the Chandrasekhar limit \citep{1983bhwd.book.....S}, NSs greater than a maximum mass limit ($\sim 2M_\odot$) do not exist \citep[also see][for a review]{2016ARA&A..54..401O}.

	\item Black hole (BH) -- BHs similarly originate from core collapse supernovae or perhaps direct collapse of very massive stars. 
	However, the mass function of the resulting BHs
	is poorly known. Before the LIGO GW observations, it was thought that BH masses 
	are
	in a narrow range around $\sim 8~M_\odot$ based on observations of $X$-ray binaries \citep{1998ApJ...499..367B}. 
	However, the LIGO GW observations have revealed the existence of more massive BHs. The 10 GW events of BBH
	mergers, found by the OI/OII runs of 
	the LIGO/VIRGO collaboration, indicate that a mass function of the BH progenitors is consistent with a Salpeter form, given as $\mathrm{d}n/\mathrm{d}M_{\rm BH}\propto M_{\rm BH}^{-2.3}$, although the constraint on the power-law slope is not tight
	\citep{2019ApJ...882L..24A}.  
	In this paper, we study how microlensing observation can be used to explore the shape of the BH mass function, and will consider several models of the BH mass function. Once the shape of BH mass function is assumed, we determine the normalization via the identity: 
	\begin{align}
	\int_{0}^{\infty}\mathrm{d}\ln M_{\rm BH}~ \phi_{\rm BH}(M_{\rm BH})=\int_{M=20}^{M=100}\mathrm{d}\ln M \frac{\mathrm{d}n_{\rm s}}{\mathrm{d}\ln M}.
	\label{eq:bh_norm}
	\end{align}
	Throughout this paper we employ $100M_\odot$ for a maximum mass scale of ZAMS progenitors of BHs. This is not an important assumption, because 
	the integration on the r.h.s. of the above equation is dominated by the lower bound, $20M_\odot$. Even if we change the upper bound to $70M_\odot$ 
	from $100M_\odot$, it changes the abundance of BHs only by 10\%.
\end{itemize}
The number conservations relating ZAMS stars to the surviving main-sequence stars and the stellar remnants give
the following ratios of the numbers between different populations for our fiducial model of ZAMS initial mass function
($\alpha_{\rm MS1}=1.3$ and $\alpha_{\rm MS2}=2.0$ in Eq.~\ref{eq:IMF})\footnote{The ratios given in this paper are slightly different from 
those in \citet{2019PhRvD..99h3503N}. The difference is from the fact that we assume $M=100M_\odot$ for a maximum mass of the ZAMS stars in this paper, 
while \citet{2019PhRvD..99h3503N} used $M=40M_\odot$.}: 
\begin{align}
\mbox{MS:WD:NS:BH}=1:0.13:0.011:0.0058.
\label{eq:number_ratio}
\end{align}
Since our MW
Galaxy consists of about $10^{11}$ stars, the ratio above means that there should be about 0.6~billion BHs in the MW in total, including both isolated systems and binary systems.
As we described above, once the normalization parameter of ZAMS mass function, $A_{\rm MS}$, is given, we can obtain the normalizations of 
the mass functions for WD, NS, and BH. This is a critical procedure for the results in this paper. 

The purpose of this paper is to discuss how 
a microlensing observation can be used to constrain 
the BH mass function. Hence we here employ a simple model for the BH mass function that is given by a handful of parameters. We consider the following power-law mass function as our default model: 
\begin{align}
\phi_{\rm BH}(M)\propto M^{1-\alpha_{\rm BH}} \hspace{1em} \mbox{for}~ M_{\rm min}\le M\le M_{\rm cut}, 
\label{eq:BH_powerlaw}
\end{align}
where $\alpha_{\rm BH}$ is a parameter to model the power-law index, $M_{\rm min}$ is a parameter to model the minimum BH mass, and $M_{\rm cut}$ 
is a parameter to model the maximum BH mass or the cut-off BH mass. 
Note that the power-law index is $1-\alpha_{\rm BH}$ instead of simply $\alpha_{\rm BH}$ since we define the mass function, $\phi_{\rm BH}$, as 
the number density per {\it logarithmic} interval of mass ($\mathrm{d}n/\mathrm{d}\ln M$).  
Our default model is given by $\alpha_{\rm BH}=2.0$, $M_{\rm min}=8~M_\odot$, 
and $M_{\rm cut}=70~M_\odot$. This model leads to $\bar{M}\simeq 19.6M_\odot$ for the average mass and $M_{0.45}\simeq13.3M_\odot$ corresponding to 
the mass point where the normalized fraction of BHs is in the $45^{\text{th}}$ percentile. The latter number might be compared to Table~4 in \citet{2020arXiv200110492W}.
The number fraction of BHs above a certain mass is $f=0.77, 0.32, 0.17, 0.097, 0.05$ or 0.02 for BHs with masses greater than 
$10, 20, 30, 40, 50$ or $60M_\odot$, respectively. Combining this with Eq.~(\ref{eq:number_ratio}),  our default model assumes 
about $5.6\times 10^{-4}(\simeq 0.0058\times 0.097)$ BHs of $M\ge 40M_\odot$ per main-sequence star.

Fig.~\ref{fig:IMF} illustrates how the mass function of ZAMS stars is related to mass functions of the stellar remnants, WD, NS and BH. 
As we described, ZAMS stars with $M_{\rm ZAMS}>1~M_\odot$ in the MW disk and bulge regions have already evolved
into their respective stellar remnants. 
Since we impose the number conservation between the number of the progenitor ZAMS stars and that of the stellar remnants, the areas of 
the correspondingly colored shaded regions are the same. For BH, the black dashed line denotes the mass function that was thought reasonable before the 
LIGO GW observation; BHs have a narrow Gaussian distribution of their masses around $7.8~M_\odot$. The black solid line shows a Salpeter-like mass function, which is consistent with the 10 BBH GW events of the LIGO/Virgo observation \citep{2019ApJ...882L..24A}; here we assume $\alpha_{\rm BH}=2.0$, $M_{\rm min}=8~M_\odot$ and $M_{\rm cut}=70~M_\odot$ for the parameters in Eq.~(\ref{eq:BH_powerlaw}). The areas under the dashed and solid black lines are the same, but these two models lead to totally different distributions of the microlensing timescales thanks to its strong dependence of the event rate on 
mass ($\mathrm{d}\Gamma/\mathrm{d}t_E\propto M^2$),  
as we will show below. 
Note that we assume there is a mass gap between NS and BH, in the range of $\sim [2,5]~M_\odot$, 
although there is a recent claim finding a BH in this mass range 
\citep{2020A&A...636A..20W}.

Furthermore, we assume a binary fraction of 0.4. 
For simplicity, we consider equal-mass binary systems: we treat 
microlensing of binary  systems
as a lens with mass 
$M_{\rm binary}=2M$ for each population. 
We do not consider binary systems that contain two objects of different masses or contain two objects of different populations
 (e.g., MS-WD system) for simplicity. Consequently we decrease the number of lens systems from the above numbers in Fig.~\ref{fig:IMF}
 by the binary fraction. So we do not include the lensing effects of caustics in a binary lens \citep[e.g. see][for the lightcurve due to a binary lens]{2020AA...633A..98W}, 
 and simply treat the effect by the increase in mass (i.e. by a point mass lens with doubled mass).
 Including the binary systems gives a slightly improved agreement between the model predictions and the OGLE data
 \citep{2017Natur.548..183M,2019PhRvD..99h3503N}. 

\subsection{A determination of the normalization of BH mass function}

To further proceed with a calculation of the microlensing event rates (Eq.~\ref{eq:dGamma/dt_E}), we need to model the spatial and velocity distributions of the lensing objects 
(MS stars and the stellar remnants) in the Galactic bulge and disk regions.
In this paper we employ the same method as in \citet{2017arXiv170102151N}, and here briefly describe the method. 

For the mass distribution of the Galactic bulge, we employ the model of spheroidal mass profile as in \citet{1992ApJ...387..181K}: 
\begin{equation}
  \rho_{\rm b}(x,y,z) =
  \begin{cases}
    1.04~M_\odot/{\rm pc}^3 
    \times {\displaystyle 10^6 \left(\frac{s}{0.482}\right)^{-1.85}}, & (s < 938~{\rm pc}) \\
    3.53~M_\odot/{\rm pc}^3~  {\displaystyle K_0\!\left(\frac{s}{667}\right)}, & (s \geq 938~{\rm pc}), \\
  \end{cases}
  \label{eq:rho_bulge}
\end{equation}
where $K_0(x)$ is the modified Bessel function of zeroth order of the second kind, $s^4 \equiv R^4$ + $(z/0.61)^4$, 
and $R^2\equiv x^2 + y^2$ (see Section~\ref{subsec:coordinates} for the definition of the coordinate system). Note the coordinates, $s$, $x$, $y$ and 
$z$ are all in units of parsec (pc). 

For the Galactic disk, we employ the exponential disk model as in \citet{1986ARA&A..24..577B}: 
\begin{align}
\rho_{\rm d}(R,z)&= 0.06~M_\odot/{\rm pc}^3\times\exp\left[-\left(\frac{R-8000}{3500}+\frac{|z|}{325}\right)\right].
\label{eq:rho_disk}
\end{align}
This model assumes that the disk has an exponential mass distribution with vertical and radial scale lengths of 325~pc and 3500~pc, respectively. 

The above Galactic bulge and disk models are based on various observations such as the luminosity functions and kinematics of stars. 
The Galactic bulge and disk models (Eqs.~\ref{eq:rho_bulge} and \ref{eq:rho_disk}) have the form given by
\begin{align}
\rho(x,y,z)&= \rho_\ast f(x,y,z),
\end{align}
where $\rho_\ast$ is the normalization constant, which has a dimension of $[M_\odot/{\rm pc}^3]$, and the function $f(x,y,z)$ is a dimension-less function that describes the spatial structure of the mass distribution. For the models of Eqs.~(\ref{eq:rho_bulge}) and (\ref{eq:rho_disk}), 
$\rho_\ast=1.04$ or $3.53~M_\odot/{\rm pc}^3$ for the Galactic bulge, while $\rho_\ast=0.06~M_\odot/{\rm pc}^3$ for the Galactic disk. 

Recalling the fact that the total stellar mass is given by the integral $\int\!\mathrm{d}\ln M~M\mathrm{d}n/\mathrm{d}\ln M$, the integrand peaks at $0.5~M_\odot$ as indicated 
by Fig.~\ref{fig:IMF}. This means that the stellar mass of the MW is dominated by MS stars around $\sim 0.5~M_\odot$. 
Assuming Eq.~(\ref{eq:IMF})
for the mass function of MS stars, we can relate the normalization of the mass function, $A_{\rm MS}$, to the above $\rho_\ast$ via
\begin{align}
\rho_\ast=\int_{0.08M_\odot}^{M_\odot}\!\!\mathrm{d}\ln M~ M\frac{\mathrm{d}n}{\mathrm{d}\ln M}.
\end{align}
Inserting Eq.~(\ref{eq:IMF}) with $\alpha_{\rm MS1}=1.3$ and $\alpha_{\rm MS2}=2.0$ into the above equation determines the normalization constant 
$A_{\rm MS}$:
\begin{align}
A_{\rm MS}\simeq \frac{\rho_\ast}{0.863M_\odot}.
\label{eq:A_MS_determination}
\end{align}
The dimension of $A_{\rm MS}$ is $[{\rm pc}^{-3}]$ (i.e. the dimension of the number density as desired). In turn, this determines the normalizations of WD, NS, and BH for an assumed form of the BH mass function. 
We further assume the same mass distribution
for each of the MS stars and the stellar remnants (WD, NS, and BH) everywhere in the disk and bulge regions. For example,
from Eqs.~(\ref{eq:phi_NS}), (\ref{eq:A_NS}) and (\ref{eq:A_MS_determination}), the spatial distribution of NS is given as 
\begin{align}
n_{\rm NS}(x,y,z; M_{\rm NS})\mathrm{d}M_{\rm NS}&\simeq \frac{0.035\rho_\ast}{0.863M_\odot}f(x,y,z)\nonumber\\
&\hspace{-3em}\times 
\frac{1}{\sqrt{2\pi}\sigma_{\rm NS}}\exp\left[-\frac{(M_{\rm NS}-\bar{M}_{\rm MS})^2}{2\sigma_{\rm NS}^2}\right]\mathrm{d}M_{\rm NS},
\end{align}
where we have used $A_{\rm NS}\simeq 0.035A_{\rm MS}$ from Eq.~(\ref{eq:A_NS}) assuming $\alpha_{\rm MS2}=2.0$.
Similarly, we can determine the spatial distributions and mass spectra at each spatial position in the MW for MS stars, WD, and BH populations
under the assumptions we employ. 

\subsection{The velocity distribution of BH} 
\label{subsec:velocity_distribution}

Now we describe the model of the velocity distribution for MS stars and the stellar remnants. We again employ the standard Galactic model for the velocity structure and assume that all the lensing objects obey the same velocity structure as that of MS stars, which are well studied by proper motion measurements \citep[e.g.][]{2020arXiv200412899Q}. BHs might have a kick velocity 
similar to that of neutron stars, which is believed to originate from asymmetric supernova explosions. However, in this paper we are most interested in massive BHs with $\ge 10~M_\odot$. 
Such massive BHs might be from a direct collapse \citep{2016Natur.534..512B} or 
a lifetime with less mass loss than that of lower mass BHs, such that
they are sufficiently massive even after a supernova explosion \citep{2001A&A...369..574V}.
Hence, for such massive BHs, we would expect a small kick velocity. In this paper we simply
assume the same velocity distribution for BHs
as that of MS stars, and will give some discussion on this issue in the conclusion section.

The two-dimensional relative velocity components $\bbv_\perp$ (Eq.~\ref{eq:v_relative}), relevant for microlensing,  arise 
from combined contributions of the Galactic rotation and the random proper motions for each of source star, lens, and observer. 
We assume that the velocity function of $\bbv_\perp$ is given by the multiplicative form, for simplicity: 
\begin{align}
f(\bbv_\perp)\mathrm{d}^2\bbv_\perp=f(v_y)f(v_z)\mathrm{d}v_y\mathrm{d}v_z.
\end{align}
In this paper, we assume a Gaussian form for each velocity component, parameterized by the mean and the width: 
\begin{align}
f(v_i)=\frac{1}{\sqrt{2\pi}\sigma_v}\exp\left[-\frac{(v_i-\bar{v})^2}{2\sigma_v^2}\right], 
\end{align}
where $\bar{v}$ is the mean and $\sigma_v$ is the rms. 

Following \citet{1995ApJ...447...53H}, we assume that 
the velocity function for a lens in the bulge  is given by
\begin{align}
&f_y: \left\{-220(1-\beta),\sqrt{1+\beta^2}100 \right\} \nonumber\\
&f_z: \left\{0,\sqrt{1+\beta^2}100 \right\},
\label{eq:fv_bulge}
\end{align}
where $\beta=d_{\rm l}/d_{\rm s}$, the first quantity in the curly brackets
denotes the mean ($\bar{v}$), and the second quantity denotes the rms ($\sigma_v$). Note that all the quantities are in units of $[{\rm km}~{\rm s}^{-1}]$.  Here we assume $100~{\rm km}~{\rm s}^{-1}$ for
the random velocity dispersion per component and $220~{\rm km}~{\rm s}^{-1}$ for the Galactic rotation (rotation velocity of an observer with respect to the Galactic center). Here we also simply assume that the $y$-direction is along the Galactic rotation direction, which is a good approximation for an observation of the Galactic bulge region as $\ell,b\approx 0~$degrees. 

For a lens in the Galactic disk, we assume that the velocity function is given by 
\begin{align}
&f_y: \left\{220\beta,\sqrt{(\kappa\delta+30)^2+(100\beta)^2}\right\}, \nonumber\\
&f_z: \left\{0,\sqrt{(\lambda\delta+30)^2+(100\beta)^2}\right\},
\label{eq:fv_disk}
\end{align}
where $\kappa=5.625\times 10^{-3}~{\rm km}~{\rm s}^{-1}~{\rm pc}^{-1}$, $\lambda=3.75\times 10^{-3}~{\rm km}~{\rm s}^{-1}~{\rm pc}^{-1}$, and 
$\delta = x - 8000~{\rm pc}$ ($x$ is in units of pc). Thus we include the spatial gradient of the 
velocity dispersion of stars against distance from the Galactic center.

\section{Results}
\label{sec:results}

\begin{figure*}
\centering
\includegraphics[width=0.9\textwidth]{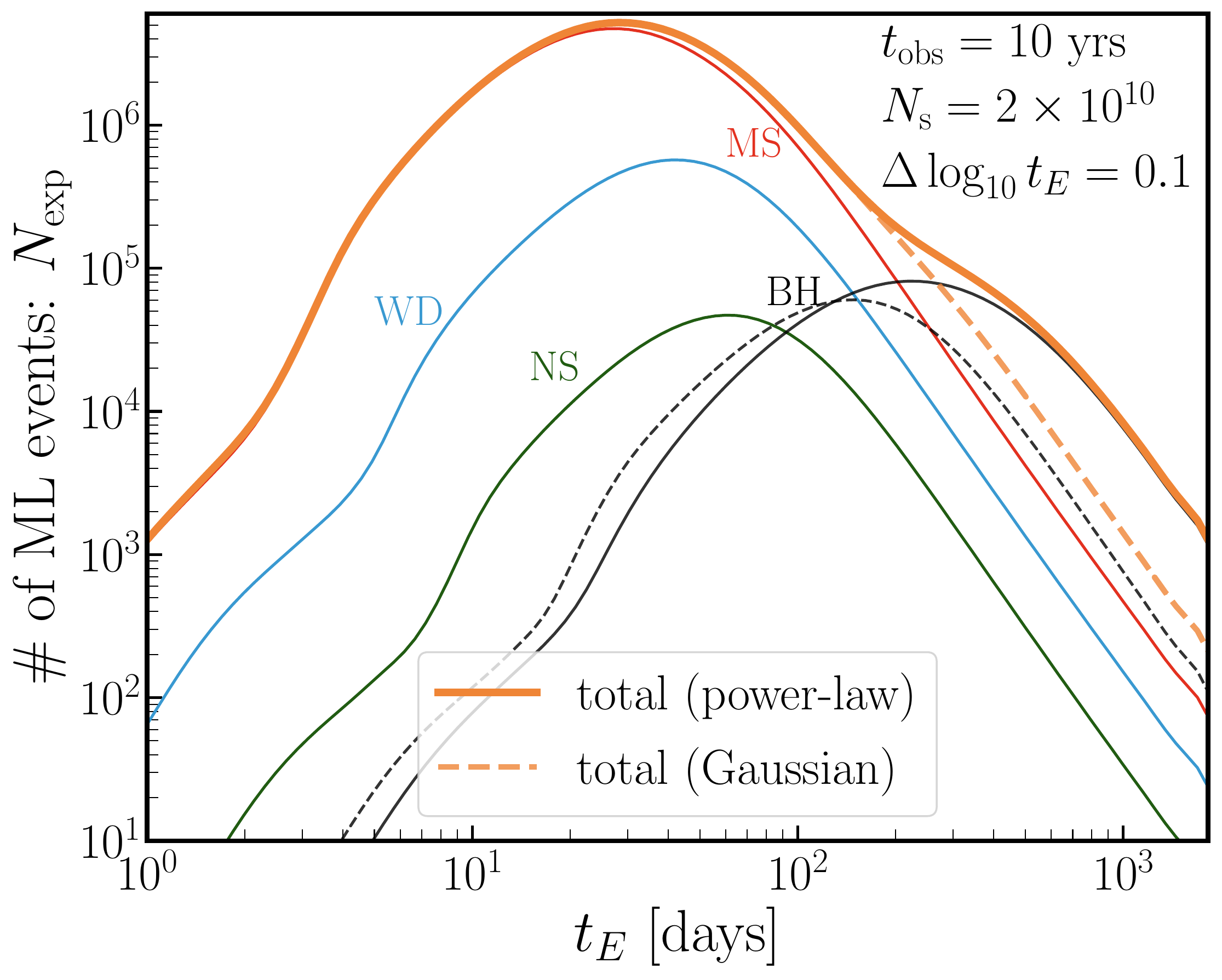}
\caption{Expected number of microlensing events in each bin of the light curve timescale ($t_E$),
for a hypothetical monitoring observation of source stars 
in the Galactic bulge region with the LSST, 
obtained by assuming the star and stellar remnant populations 
and the standard model of Galactic structures (spatial and velocity distributions). 
Here we assume $t_{\rm obs}=10~{\rm yrs}$ for the duration of observation, $N_s= 2 \times 10^{10}$
for the number of source stars, and 
$\Delta\log_{10}t_E=0.1$ (10 bins in one decade of $t_E$) corresponding to $\Delta\ln t_E\simeq 0.23$. We assume 60\% efficiency 
for all the timescale microlensing events (see text for details), but note that the number events for low $t_E$ would be overestimated (an actual efficiency would be lower depending on details of the cadence).
The thin red, blue and green lines are the contributions from main-sequence (MS) stars, white dwarfs (WD), and neutron stars (NS), respectively, assuming their mass functions in Fig.~\ref{fig:IMF}. The black dashed and solid lines are the contributions from black holes (BH) assuming the Gaussian and power-law mass functions in Fig.~\ref{fig:IMF}, respectively. 
The top, thick
solid and dashed lines are the total contributions for the two models of black holes.
\label{fig:num_exp}}
\end{figure*}

\begin{figure}
    \centering
    \includegraphics[width=0.5\textwidth]{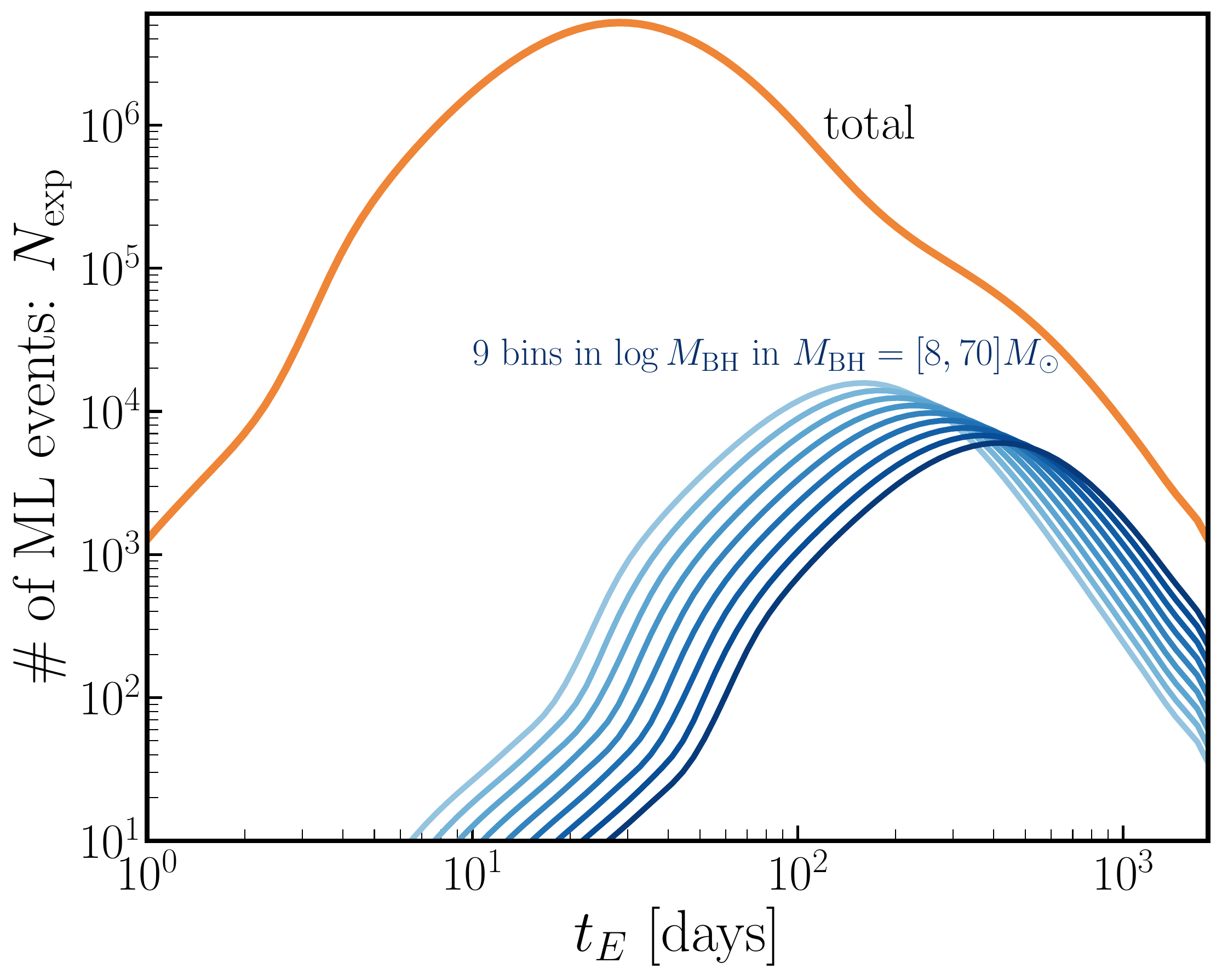}
    \caption{Similar to the previous plot, but blue lines show differential contributions of black holes, divided into different mass bins, to the total 
     number of microlensing events. The top, thick
     line is the same as in Fig.~\ref{fig:num_exp}, showing
     the total number obtained assuming the power-law mass function of black holes in Fig.~\ref{fig:IMF}. The blue lines show the contributions for each of 9 black hole subsamples that are evenly spaced in the logarithmic space 
     of black hole mass in the range $M_{\rm BH}=[8,70]~M_\odot$. 
     From 
     light to dark lines, the central mass of each bin corresponds to $M_{\rm BH}/M_\odot\simeq 9$, 11, 15, 
     19, 24, 30, 38, 49, and $62$, respectively.   
     }
    \label{fig:events_mbin}
\end{figure}

\begin{figure}
    \centering
    \includegraphics[width=0.5\textwidth]{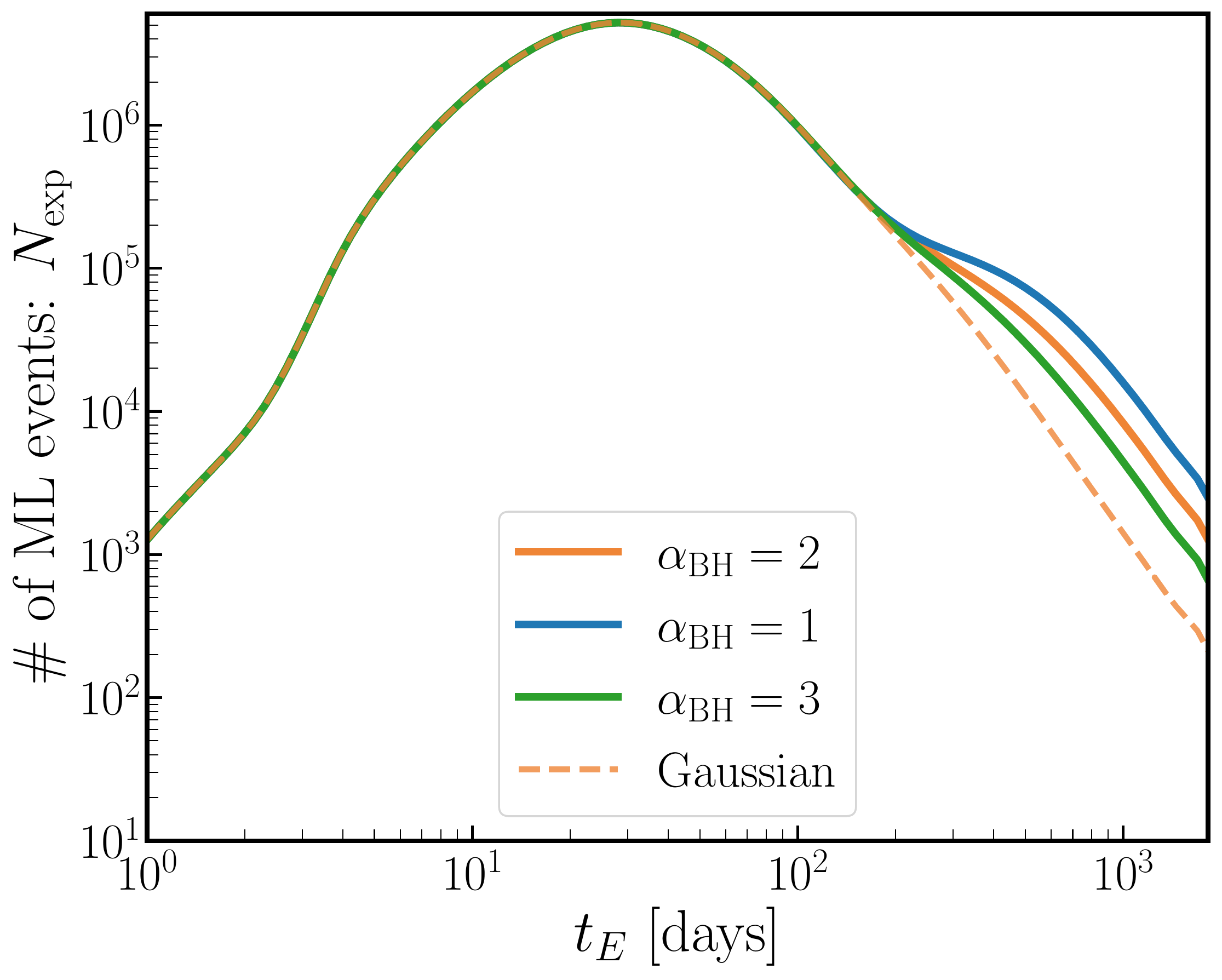}
    \caption{Dependence of the microlensing events on the mass slope of BH mass function. The orange solid and dashed lines are the same 
    as those in Fig~\ref{fig:num_exp}. The solid line is 
    our default model, where we assume a Salpeter-like slope of $\alpha_{\rm BH}=2$ for the mass slope 
    of BH mass function, defined as $\mathrm{d}n_{\rm BH}/\mathrm{d}\ln M\propto M^{1-\alpha_{\rm BH}}$.
    The blue and green lines are the results for $\alpha_{\rm BH}=1$ and 3, respectively. We impose
    the number conservation between ZAMS massive stars and BHs; 
    the number of BHs in the Galactic bulge and disk regions are 
    the same for all the models. For comparison, the thick, dashed
    line, the same as in Fig.~\ref{fig:num_exp}, 
    shows the model prediction for a model of Gaussian BH mass function
    in Fig.~\ref{fig:IMF}. }
    \label{fig:events_vary_alpha}
\end{figure}

\begin{figure}
    \centering
    \includegraphics[width=0.5\textwidth]{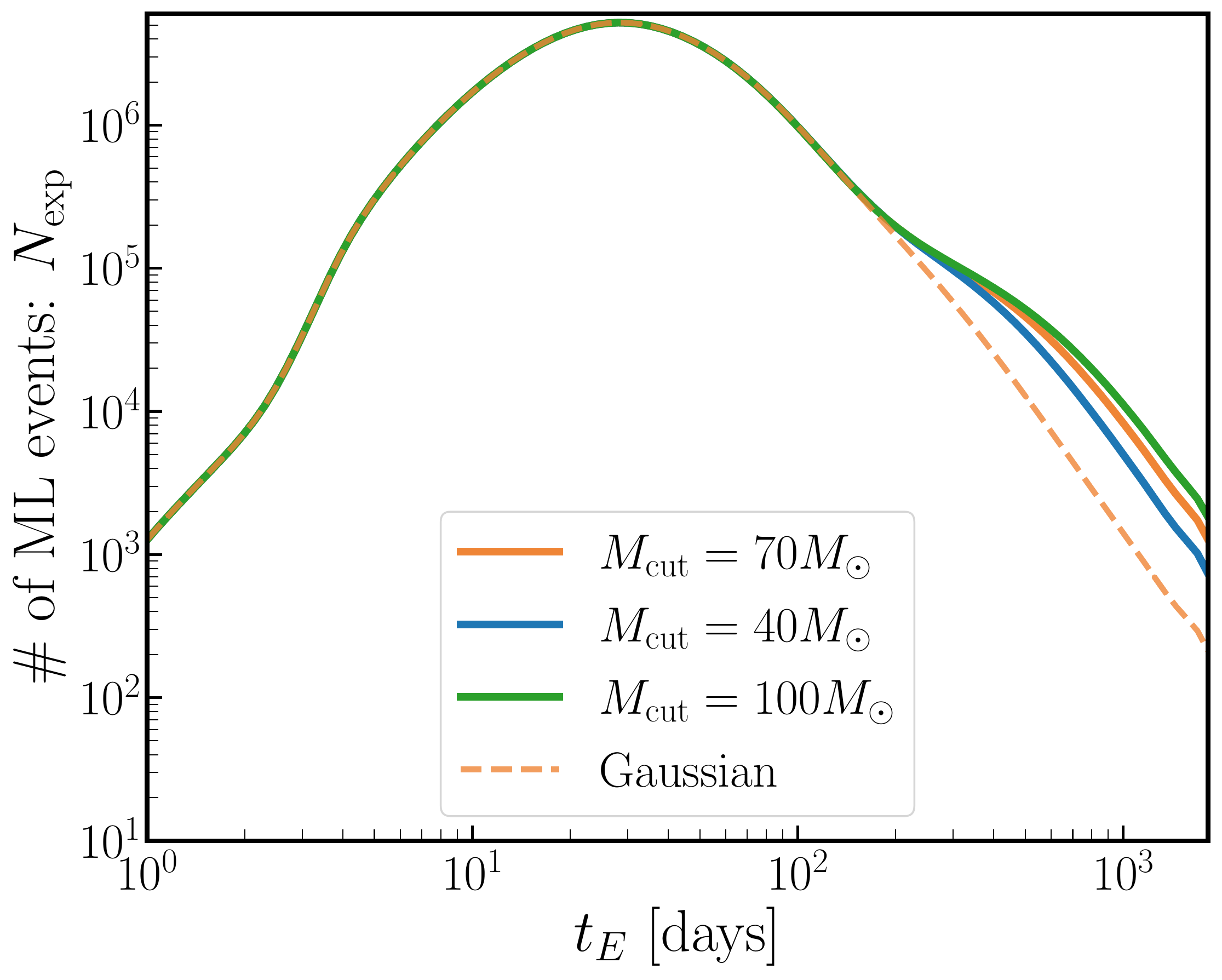}
\caption{Similar to the previous figure, but dependence of the microlensing events on the maximum mass cut of BH in the Salpeter-like mass function ($\alpha_{\rm BH}=2$). While we assume 
    $M_{\rm max}=70M_\odot$ for a maximum mass of BHs in our default model (orange line), the green and blue lines 
    show the results if we assume $M_{\rm max}=40$ or 100$M_\odot$, respectively. Again we impose
    the number conservation: all the results have 
    the same number of BHs.}
    \label{fig:events_vary_M_cut}
\end{figure}

\begin{figure}
    \centering
    \includegraphics[width=0.5\textwidth]{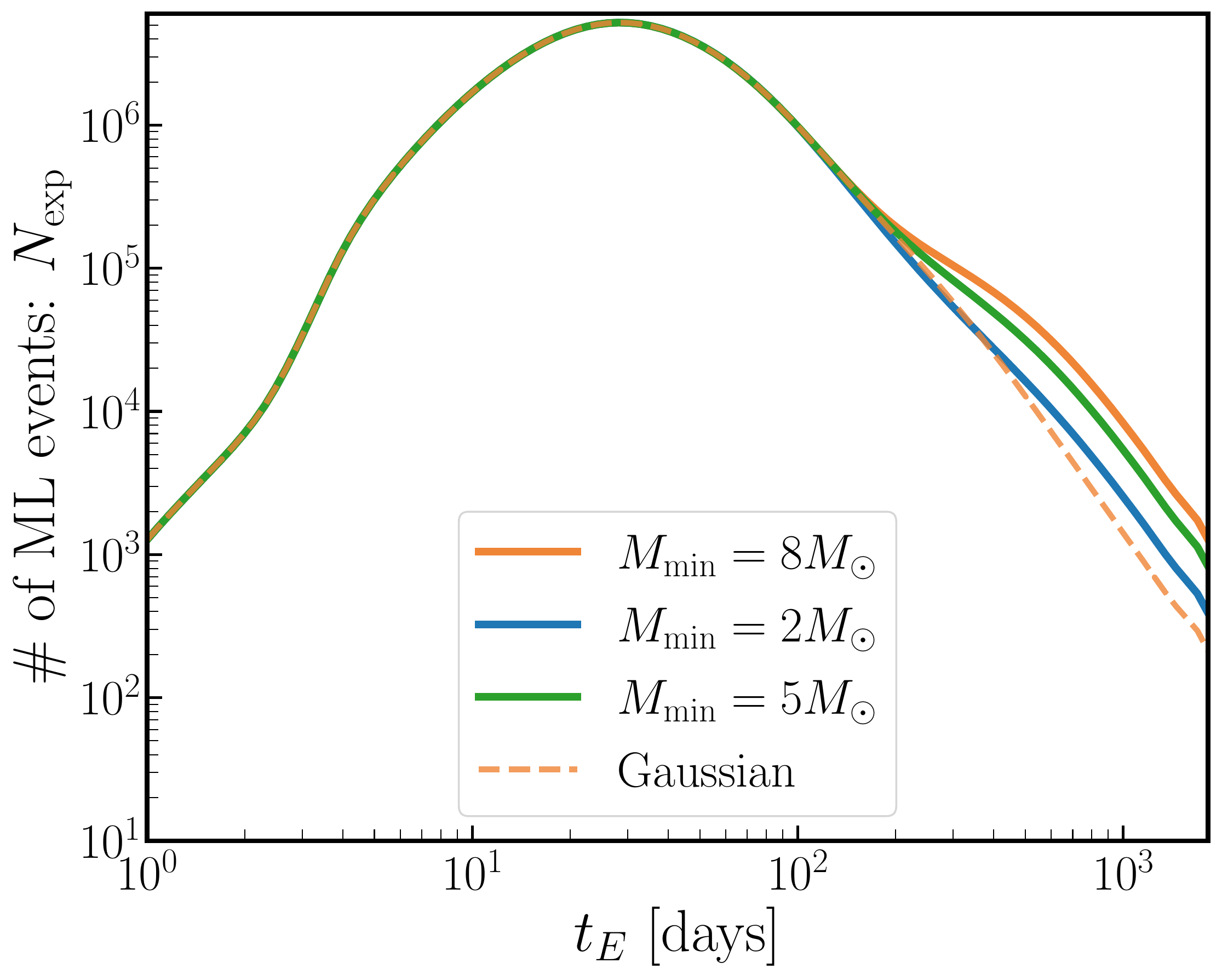}
\caption{Similar to the previous figure, but in this figure we probe the microlensing dependence on BHs in the nominal mass gap ($[2,8]M_\odot$)
by varying the minimum mass of BHs. In our fiducial model, we assume 
    $M_{\rm min}=8M_\odot$ for a minimum mass of BHs (orange line). The green and blue lines 
    are if we assume $M_{\rm min}=2$ or 5$M_\odot$, respectively. Again, all the results have 
    the same number of BHs.}
    \label{fig:events_vary_M_min}
\end{figure}

\begin{figure}
    \centering
    \includegraphics[width=0.47\textwidth]{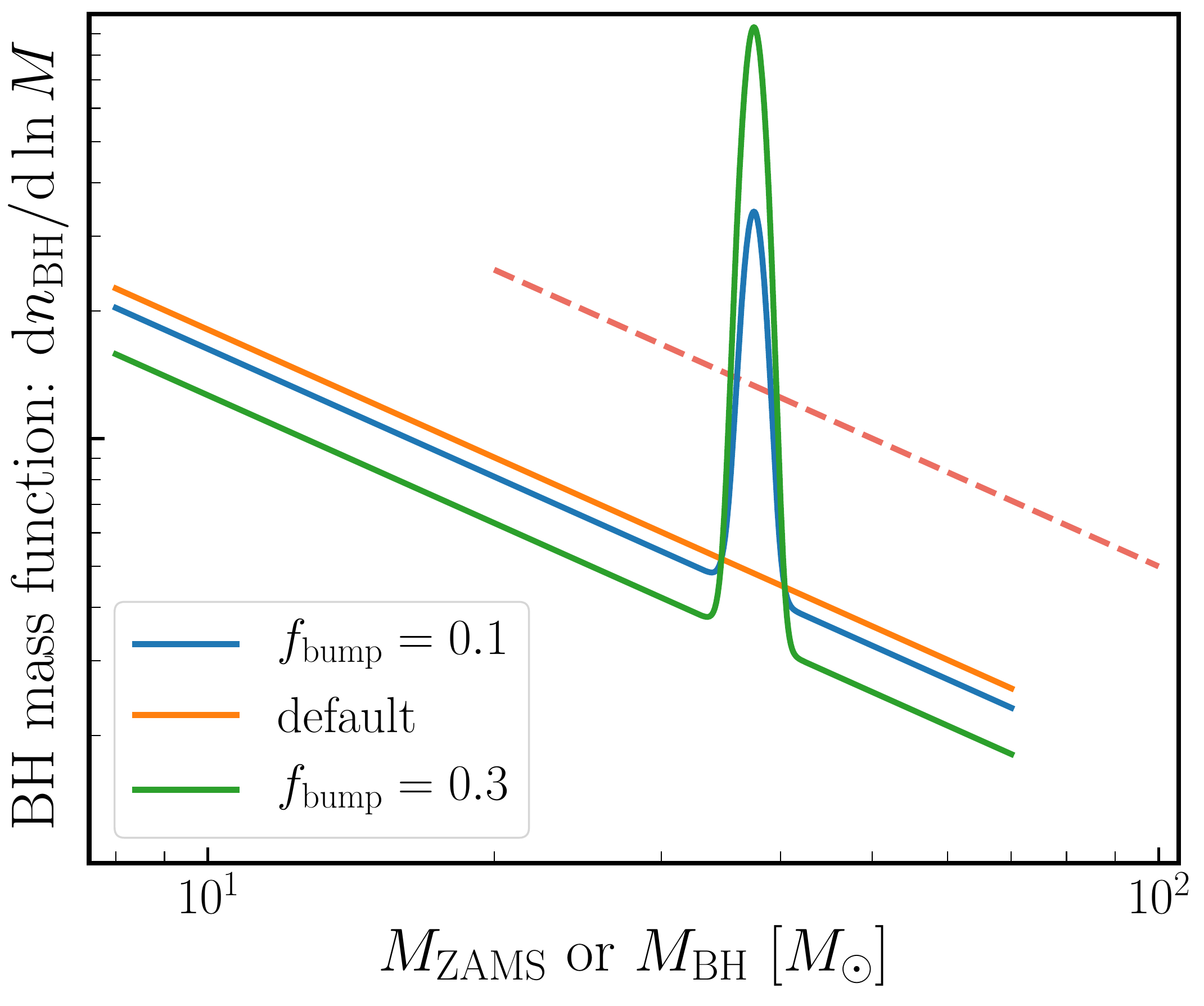}
    \includegraphics[width=0.5\textwidth]{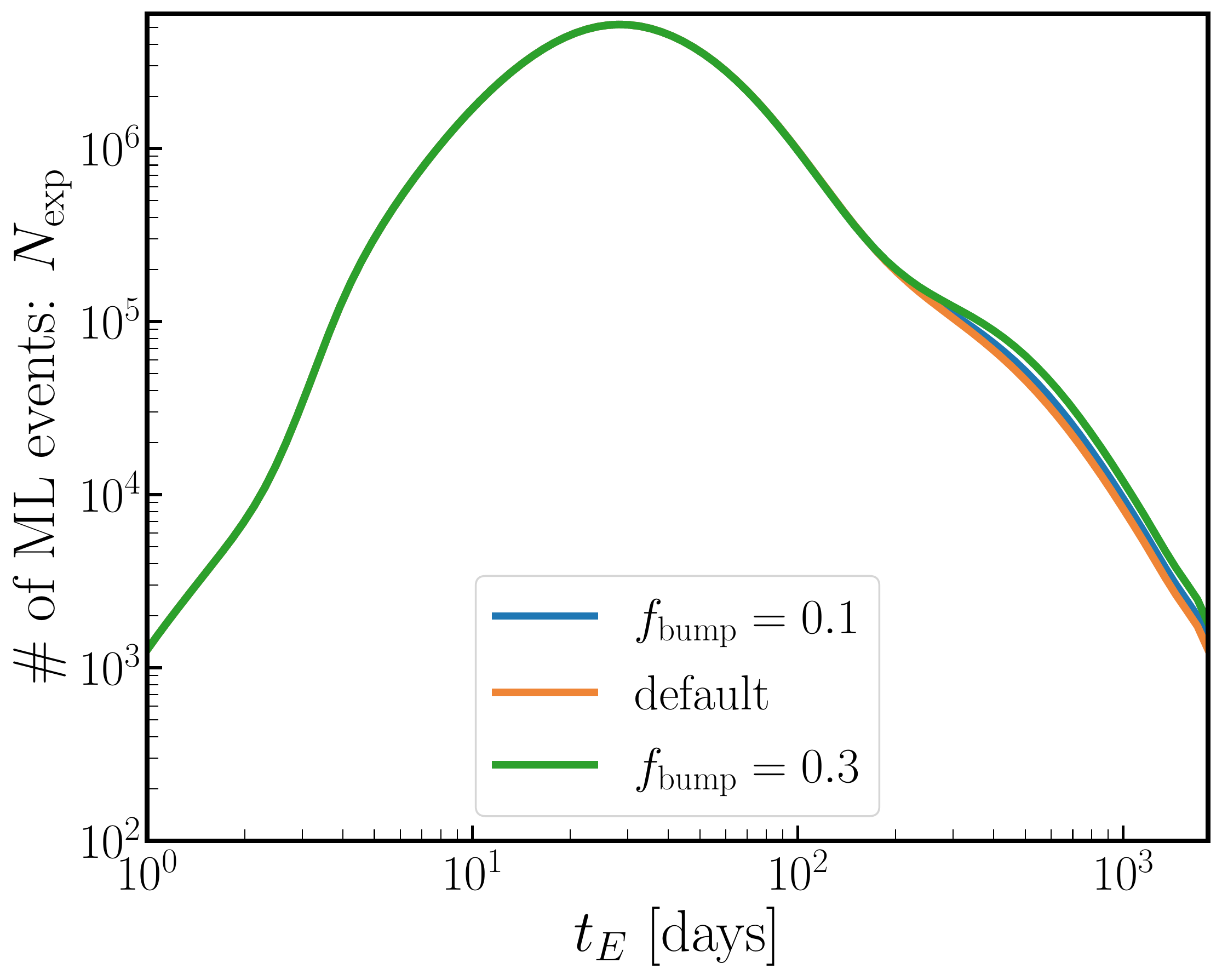}
\caption{{\it Upper panel}: The BH mass function 
that is modeled by a sum of the power-law contribution and the Gaussian component, which is motivated by the recent study in \citet{2020arXiv200110492W}. For the Gaussian component, we consider a Gaussian with mean $M_{\rm bump}=37.5M_\odot$ and 
width $\sigma=1M_\odot$, while we consider $\alpha_{\rm BH}=2$ for the slope of the power-law component. We parametrize the mass function by a fraction of the Gaussian component to the total number of BH: we consider $f_{\rm bump}=0$ (default), $0.1$, or $0.3$. 
All the models have the same number of BHs. For comparison, the dashed line shows the mass function of ZAMS progenitors of BHs. 
 {\it Lower panel}: Expected number of microlensing events for the three cases of BH mass function.}
    \label{fig:events_powerlaw_bump}
\end{figure}

We are now in a position to compute the event rate for each population (MS, WD, NS, or BH) by plugging the number density and velocity distributions of each population we have obtained up to the preceding section into 
Eq.~(\ref{eq:dGamma/dt_E}). 
In this section, we show the main results of this paper. 

\subsection{The impact of LIGO-GW mass scale BHs on the microlensing events}
\label{sec:Nexp}

The expected number of microlensing events in a given range of the $i$-th light curve timescale bin, $t_E=[t_{E,i}-\Delta t_{E,i}/2,
t_{E,i}+\Delta t_{E,i}/2]$, for a given monitoring observation of the Galactic bulge region is given as
\begin{align}
N_{\rm exp}&=t_{\rm obs}N_{\rm s}\int_{t_{E,i}-\Delta\!t_{E,i}/2}^{t_{E,i}+\Delta\!t_{E,i}/2}\!\mathrm{d}t_E~
\frac{\mathrm{d}\Gamma}{\mathrm{d}t_{E}}\epsilon(t_E)\nonumber\\
&\simeq t_{\rm obs}N_{\rm s}\left.\frac{\mathrm{d}\Gamma}{\mathrm{d} t_E}\right|_{t_{E,i}}t_{E,i}\times \Delta\ln t_{E}\times \epsilon(t_{E,i}),
\label{eq:N_exp}
\end{align}
where $t_{\rm obs}$ is the duration of the monitoring observation, $N_{\rm s}$ is the number of source stars, 
$\Delta\ln t_E$ is the bin width in the logarithmic timescale intervals, 
and $\epsilon(t_{E})$ is a 
detection efficiency quantifying the probability that a microlensing event of timescale $t_E$ is detected by 
a given
observation. 
In the following 
we consider logarithmically evenly-spaced bins of $t_E$, so $\Delta \ln t_E={\rm constant}$.

In this paper, we consider a hypothetical monitoring observation with LSST over 10~years (i.e. $t_{\rm obs}=10~$yrs).
Thanks to the anticipated excellent image quality, the large aperture, and the wide field-of-view, LSST is expected to produce an ideal dataset for exploring microlensing events. Following \citet{2019ApJ...873..111I}
we assume that such a dedicated LSST observation allows for monitoring observation of more than
billions of stars in the Galactic bulge region: that is, we assume $N_{\rm s}= 2 \times 10^{10}$.
This number is contrasted with that for the OGLE experiment, which has been using about $4.88\times 10^7$ source stars with the 1.3~m dedicated telescope \citep{2017Natur.548..183M}. Hence we think $N_{\rm s}= 2 \times 10^{10}$
for LSST 
is reasonable, but needs a more careful study based on actual data, 
e.g. data from the Dark Energy Camera as a pilot observation. As is obvious from the above equation, if we 
use a smaller number of source 
stars than what we assume, the number of events simply decreases by the ratio factor. 
A blending of multiple stars in the source plane will also have an effect on how many events we are able to observe
\citep{2019PhRvD..99h3503N}. 
In the presence of blending effects, 
the observed amplification is modified by $A_{\rm obs} = (A-1)f_s +1$ where $A$ is the true amplification and $f_s$ is the normalized source flux such that $f_s$ = 1 in the absence of blending. As discussed in \cite{2015ApJS..216...12W},
the maximum efficiency with blending in OGLE-III is 60\%.
We would need a dedicated study to determine the efficiency function of LSST; e.g. 
we could use simulated images or inject simulated 
images of microlensing light curves into actual data, and then study the probability of recovering the injected events as a function of various observation conditions. This is beyond the scope of this paper. Instead, we will below discuss how efficiency depends on the observation cadence. 
As discussed in section \ref{sec:cadence}, in particular, we expect the efficiency to be lower for the events with shorter Einstein crossing times.
For the moment, we will assume $\epsilon$ = 0.6 since the effects of blending are thought to be similar to OGLE-III and therefore have a similar efficiency. 


In Fig.~\ref{fig:num_exp} we show the expected number of microlensing events as a function of the light curve timescale $t_{E}$, expected for 
a hypothetical 10-year monitoring observation of $2 \times 10^{10}$
stars in the Galactic bulge with LSST. Here we 
used the model described in Section~\ref{sec:mirolensing}. 
First of all, the figure clearly exhibits that LSST would enable 
one to find 
millions of microlensing events in each of timescale bins,
showing the power 
of LSST, if $2 \times 10^{10}$
source stars are used for the microlensing search.   
The different lines show the results for different populations of MS, WD, NS, and 
BH, assuming their mass functions in Fig.~\ref{fig:IMF}. Most notably, the black dashed and solid lines for 
the Gaussian and power-law mass function of BHs, respectively, give substantially different numbers of microlensing events, even though the two models have the same 
number of BHs. A power-law model for BHs predicts that BHs are a dominant source of 
microlensing events in each bin of 
$t_E\gtrsim 100~$days, even if the number of BHs is only 0.6\% of the number of MS stars in our model (Eq.~\ref{eq:number_ratio}). 
To be more precise, an LSST observation 
would allow us to find about $6.1\times 10^5$
BH events. 
A Gaussian BH model leads to a factor of about 3 fewer events for microlensing events with $t_E>200~{\rm days}$, where
the BH microlensing is a dominant source of the total events. 
For events at $t_E\gtrsim 10^3$~days, most 
events are from BHs with masses greater than  $20M_\odot$.
This boosted number of microlensing events at such long timescales is due to the mass boost in the microlensing event rate for such long-timescale 
events 
($\Gamma\propto M^2$), as explained 
around Eq.~(\ref{eq:gamma_mass_scaling}). Also note that all lines of each population show an asymptotic behavior of $N_{\rm exp}\propto t_E^{-3}$ 
in large $t_E$ bins
in this logarithmically-spaced binning, again as explained by Eq.~(\ref{eq:gamma_mass_scaling}). 
This is a 
consequence of the $t_E$-distributions if all the lens populations follow the same (or similar) spatial and velocity distributions. Thus we conclude that, if BHs have an underlying power-law mass distribution extending to $\gtrsim 20M_\odot$, 
a measurement of microlensing $t_E$-distribution allows for a direct test of the existence of such heavy BHs in the MW Galaxy, 
in the statistical sense. Since these events have a long timescale light curve, it would also be easier to make a follow-up observation of individual events, which would help to discriminate secure candidates of BH microlensing on an individual basis. 
Note that the numerical 
integration of 
Eq.~(\ref{eq:dGamma/dt_E}) fully converges for $t_E \gtrsim 10^2$~days, which is our timescale of interest.

Blue lines in Fig.~\ref{fig:events_mbin} show the differential contributions of BHs, divided into different mass bins, to the total event number.
All the lines have an asymptotic behavior of the $t_E$-distribution as $N_{\rm exp}\propto t_E^{-3}$. The figure clearly shows that the microlensing 
events at longer timescales are dominated by heavier BHs, as expected. However, each line displays a wide $t_E$-distribution, spanning two orders of magnitude in $t_E$ ($x$-axis), 
with the range of event number varying over one order of magnitude ($y$-axis).
BHs with $M\gtrsim 30M_\odot$ give about $2 \times 10^5$ events, about one third of all BH events. 

In Fig.~\ref{fig:events_vary_alpha} we study how varying the slope of the BH mass function, parametrized by $\mathrm{d}n_{\rm BH}/\mathrm{d}\ln M_{\rm BH}\propto M_{\rm BH}^{1-\alpha_{\rm BH}}$, alters the predictions of microlensing events. Our default model is a Salpeter-like mass function, given by $\alpha_{\rm BH}=2$. Note that all the models satisfy the number conservation between ZAMS massive stars and the BH remnants according to our method 
(Section~\ref{sec:BH_massfunction}): that is, all the models have the same number of BHs. As is obvious from the figure, increasing the relative populations of heavier BHs leads to a boost in the number of microlensing events at longer timescales. Among the three cases, the model with the shallowest slope, $\alpha_{\rm BH}=1$, leads to
the largest population of such long-timescale events. 

In Fig.~\ref{fig:events_vary_M_cut} we study the model predictions for differing choices of the maximum mass cut in the BH mass function: 
$M_{\rm cut}=40$, 70, or $100M_\odot$, respectively, where $M_{\rm cut}=70M_\odot$ is our default model. Similarly to other figures, in this figure, the existence of heavier BHs leads to a larger number of events at longer timescales. 

In light of the recent observation by LIGO of an object in the mass gap between black holes and neutron stars \citep{LIGO:20a, LIGO:20b} and claims outside of gravitational wave detections \citep{2019Sci...366..637T, 2020A&A...636A..20W}, we vary the lower limit of the LIGO black-hole mass function. In Fig.~\ref{fig:events_vary_M_min} we vary the lower limit of the BH mass function: $M_{\text{min}}$ = 2, 5, and 8 $M_{\odot}$ where $M_{\rm min} = 8 M_{\odot}$ is our default model. Again, heavier BHs leads to more events at longer timescales. Hence the existence of lower mass BHs, as in the case of $M_{\rm min}=2M_\odot$, leads to fewer events of longer timescales, because of the lowered normalization of the mass function for heavier BHs.

Moreover, motivated by the recent work \citep{2020arXiv200110492W} that claims that BH mass might have a preferred mass scale 
at birth as a consequence of the stellar evolution, 
we also consider a contribution to microlensing arising from 
the secondary population of BHs, modeled by a Gaussian form with $M_{\rm bump}=37.5M_\odot$ and width $\sigma=1~M_\odot$.
The preferred mass scale roughly corresponds to the mass scale of binary BH systems for the first LIGO GW event, GW150914 \citep{2016PhRvL.116f1102A}.
In the upper panel of Fig.~\ref{fig:events_powerlaw_bump}, we show the BH mass function 
which we model 
as a sum of the power-law and 
Gaussian contributions. We parametrize this model by a fraction of the Gaussian component, $f_{\rm bump}$,  
to the total number of BHs: here we consider $f_{\rm bump}=0.1$ or 0.3. 
The case of $f_{\rm bump}=0.1$ is intended to roughly reproduce 
Fig.~6 in \citep{2020arXiv200110492W}. Note that Fig.~6 in their paper plots $\mathrm{d}n_{\rm BH}/\mathrm{d}M$, instead of 
$\mathrm{d}n_{\rm BH}/\mathrm{d}\ln M$. This is a toy model, but the purpose is to study the impact of a specific feature in the BH mass function on the microlensing event rate. 
The lower panel of Fig.~\ref{fig:events_powerlaw_bump} shows the expected number of microlensing events. Even if the BH mass function has such a sharp feature around the particular mass scale, it does not imprint a corresponding feature in 
the $t_E$-distribution of microlensing events, as explained by Fig.~\ref{fig:events_mbin}. Therefore, we conclude that it is difficult to explore a narrow feature of the underlying BH mass function from the $t_E$-distribution of microlensing events in the statistical sense. Nevertheless, if we can make a follow-up study of individual secure events of BH microlensing, e.g. microlensing parallax and astrometric signal
\citep{2000ApJ...542..785G}, it would be possible to reconstruct the mass distribution of BHs.

\subsection{Evaluating LSST Cadences}
\label{sec:cadence}

\begin{figure}
    \centering
    \includegraphics[width=0.5\textwidth]{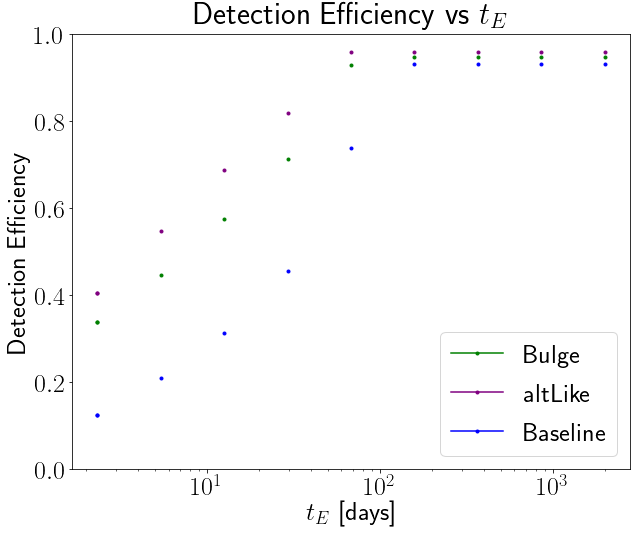}
    \caption{Detection efficiency of the three LSST cadences as a function of $t_E$ [days]. At low $t_E$, the altLike cadence does better than the bulge cadence which does better than the baseline cadence. However, at high $t_E$, all three are at approximately the same detection efficiency. They are not at 100\% due to edge effects.}
    \label{fig:efficiency}
\end{figure}

Here we study how a monitoring observation of the Galactic bulge with LSST 
allows
us to explore microlensing events. To do this, 
we evaluate the performance of the three cadences considered in the LSST collaboration: 
the ``baseline'' cadence\footnote{baseline\_v1.3\_10yrs}, which is an algorithmically determined cadence with a few deep drilling spots; the ``bulge'' focused cadence\footnote{bulges\_cadence\_bulge\_wfdv1.3\_10yrs}, which is a similarly algorithmically determined cadence with a focus on the Galactic bulge; and the ``altLike'' cadence\footnote{altLike\_v1.3\_10yrs}, which is a deterministic cadence that scans the meridian of the sky as it passes overhead. The simulated cadences are from the OpSim database \citep{2016SPIE.9910E..13D,2016SPIE.9911E..25R} made by the Rubin Observatory team, which take into account historical weather conditions and other variables. 
We extract cadence information from eight fields around the bulge (270$^{\circ}, -30^{\circ}$) each ranging 3.5$^{\circ}$ in RA and Dec around the center of the field of view to correspond to the LSST
field of view. The center of each field of view is 4$^{\circ}$ apart so they do not overlap. 

To evaluate the cadences, we simulate 10,800 light curves 
with various $t_E$, $t_0$, and $u_{\rm min}$. 
We choose ten values of $t_E$ logarithmically spaced between 0 and 2000 days; the upper bound is due to the event rate dropping off at around this value 
(see Fig.~\ref{fig:num_exp}). 
We choose 120 values of $t_0$ 
spaced evenly for the 10 year survey, so there is one month between each value. We choose ten values of $u_{\rm min}$, evenly spaced from 0 to 1. We then sample these light curves 
with the simulated cadences.

We consider a microlensing event 
``detected'' 
with A $\geq$ 1.1 and at least four observations on either side of the peak of the simulated light curve. This ensures that it is
observed at least two nights before and after the 
magnification
peak since often two observations will be taken on one night.

By plotting the detection efficiency (see Fig.~\ref{fig:efficiency})
as a function of $t_E$, we can see that the ``altLike'' cadence 
achieves
the highest efficiency,
the bulge cadence achieves the second highest, and the baseline cadence achieves the lowest. At high Einstein crossing times (over 100 days), there is no 
significant
difference between the cadences. At maximum efficiency they detect nearly all the events. It is likely that the cadences do not detect all of the events due to some of the events having their peak extremely close to when the cadences begin or end their observations, so the telescope is unable to make four observations before and after the peak for those events. Note that this evaluation does not include the presence of parallax, which is a useful quantity for identifying black holes (see section \ref{sec:conclusion}).

\section{Discussion and Conclusion}
\label{sec:conclusion}

In this paper we have studied how microlensing observation can be used to explore a population of BHs that exist in the Galactic bulge and disk regions, motived by the LIGO/Virgo GW events due to binary black hole systems that indicate a population of heavier BHs 
with masses $\gtrsim 30~M_\odot$ than  the
previously anticipated mass scale of $\sim 8M_\odot$. We showed that, if BHs have a Salpeter-like mass function extending 
up to $70M_\odot$ in our default model, the BH population yields a dominant source of microlensing events at longer timescales, 
$t_E\gtrsim 100~{\rm days}$ (Fig.~\ref{fig:num_exp}). 
We showed that a tiny population of such heavier BHs could alter the $t_E$-distribution of microlensing events at such long timescales
due to the boost of microlensing event rates given by $M^2$ for a fixed $t_E$ (Figs.~\ref{fig:events_mbin}--\ref{fig:events_vary_M_cut}). 
In our fiducial model, we assume about $6\times 10^{-4}$ BHs of $\ge 40M_\odot$ per main-sequence star. 
To find this result we assume the number conservation between BHs and the massive progenitors of zero-age main sequence stars
and that the BH population follows the 
same spatial and velocity structures as those of the surviving main sequence stars in the Galactic bulge and disk regions. 
If LSST 
carries out a suitable cadence observation to monitor $2 \times 10^{10}$
source stars towards the Galactic bulge over 10~years, it could 
detect 
about $6\times 10^5$
BH microlensing events thanks to its unique capability.
However, the Galactic bulge region usually has a large dust extinction, so it requires a more careful study of how an optical telescope such as LSST can carry out an efficient microlensing observation in the bulge region. 
Our results might be compared to \cite{2020ApJ...889...31L}, which shows 
different event rates from our results. The difference is mainly from the different BH mass function; \cite{2020ApJ...889...31L} employed the BH mass function with maximum mass cut around $M\sim 16M_\odot$, and did not include a population of the heavier BHs in the event rate evaluation. 

The origin of LIGO/Virgo BHs is poorly understood. As we showed, microlensing can be a powerful tool to explore the nature of the BH population in the MW. 
Here, microlensing is complementary to the statistical method using the GW BBH events, which are sensitive to close-orbit and heavier BHs due to the dependences of LIGO/Virgo GW sensitivities on properties of BBH systems. On the other hand, microlensing is sensitive to both isolated BHs and wide-orbit BBH systems. If a characteristic signature in the microlensing light curve due to binary systems is detected, the microlensing method could 
provide useful information on the binary fraction of BHs. Furthermore combining the microlensing constraints and the LIGO BBH population studies would give a coherent picture of the origin of the BH population. 

A major assumption in our analysis is that the velocity distribution of BHs is the same as that of the main sequence stars, 
which has been well studied. BHs 
may have a large kick velocity due to anisotropic supernova explosions, so they 
may have faster velocities on average than main sequence stars do. 
A faster-moving, heavier lens would 
produce a similar 
microlensing timescale to that of
a slower-moving, lighter lens. Thus an observation of the light curve alone would cause degeneracies in parameters of individual lenses.
In addition, if BHs have faster velocities on average than those of main-sequence stars, 
it would weaken the shoulder-like feature of the $t_E$-distribution in 
Fig.~\ref{fig:num_exp} (the $t_E$-distribution of BHs 
would shift left horizontally in the figure). To overcome this obstacle, a detailed follow-up observation of individual secure candidates of BH microlensing would be useful. For example, we can explore the microlensing parallax and astrometric signal
\citep{2000ApJ...542..785G} \citep[also see][]{2008Sci...319..927G,2010ApJ...713..837B} to disentangle the parameter degeneracies.
Such parallax signals have been measured successfully \citep{2005ApJ...633..914P, 2016MNRAS.458.3012W}, though the parallax is not enough to completely break the degeneracy. Through astrometric microlensing, one can solve for the Einstein radius and therefore the mass of the lens \citep{2000ApJ...534..213D, 2018MNRAS.476.2013R,2016ApJ...830...41L}.
Measuring the parallax
would be possible via LSST itself through the year-timescale monitoring observation due to the orbital motion of the Earth and/or if the Roman Space Telescope jointly
monitored the light curve of the same object at the same time; the upcoming Gaia NIR mission might also be a helpful 
monitor \citep{2019arXiv190712535H}. 
Combining the light curve
with proper motion measurements, pre- or post-microlensing observation would also be useful. 
For a BH lens system, one can only observe the source star, so one could disentangle properties of the lens-source system provided the combined information of proper motion and source flux \citep{2020A&A...636A..20W}, if the source star is resolved. 
The Japan-led JASMINE satellite project\footnote{\url{https://www.nao.ac.jp/en/}} \citep{2012SPIE.8442E..47K} 
would give 
useful information on the proper motion, if the observation region has an overlap with the LSST microlensing observation region. The European Extremely Large Telescope\footnote{\url{https://www.eso.org/sci/facilities/eelt/}}
would also be useful thanks to the superb angular resolution by adaptive optics.
These are all interesting possibilities
worth further 
investigation to be
presented elsewhere.

\acknowledgments
We would like to thank Christopher~Stubbs for his extensive guidance and support, particularly with regard to the LSST cadence evaluation. 
MT would also like to thank Shude~Mao for useful inputs on the earlier version of this paper. 
N. S. A. appreciates the warmth and hospitality of Kavli IPMU, where this work was initiated. 
We would like to thank Zoltan~Haiman, Hiroko~Niikura, Charles~Alcock, Toshiki~Kurita, Sunao~Sugiyama, and Satoshi~Toki for their useful discussion. 
We would like to thank the reviewer for the helpful comments.
This work was supported in part by the World Premier International
Research Center Initiative (WPI Initiative), MEXT, Japan, Reischauer Institute of Japanese Studies at Harvard University, and JSPS
KAKENHI Grant Numbers 
JP15H05887, JP15H05893, JP15H05896, JP15K21733, and JP19H00677.

\bibliographystyle{aasjournal}
\bibliography{refs}

\end{document}